\documentclass[journal,12pt,onecolumn,]{IEEEtran}

\IEEEoverridecommandlockouts                              

\usepackage{graphicx}
\usepackage{amsmath}
\usepackage{amssymb}
\usepackage{dsfont}
\usepackage{xcolor}
\usepackage{upgreek}
\usepackage{enumerate}
\usepackage{tikz}
\usepackage{amsfonts}
\usepackage{textcomp}
\usepackage[normalem]{ulem}
\usepackage{setspace}
\usepackage{comment}
\usepackage{relsize}
\usepackage{subfig}
\usepackage{cite}
\usepackage{circuitikz}

\usetikzlibrary{shapes,arrows}
\usepackage{verbatim}
\usetikzlibrary{calc}

\usepackage{arydshln,leftidx,mathtools}
\usepackage{caption}
\usepackage{float}
\usepackage{bm}
\usepackage{mathrsfs}

\newtheorem{theorem}{Theorem}[section]
\newtheorem{lemma}{Lemma}
\newtheorem{remark}{Remark}[section]

\newtheorem{defn}{Definition}[section]

\newtheorem{coro}{Corollary}[section]

\newtheorem{proposition}{Proposition}[section]

\newcommand{\norm}[1]{\left\lVert#1\right\rVert}
\newcommand{\inv}{^{\raisebox{.2ex}{$\scriptscriptstyle \!-\!1$}}}

\newcommand{\stbt}[4]{\left[\begin{smallmatrix} #1&#2\\#3&#4\end{smallmatrix}\right]}

\usepackage{hyperref}
\hypersetup{colorlinks = true,
            linkcolor = red,
            urlcolor  = blue,
            citecolor = blue,
            anchorcolor = blue}

\title {
Phase Preservation of $N$-Port  Networks under \\
General Connections
\thanks{This work was supported in part by the Research Grants Council of Hong Kong Special Administrative Region, China, under the General Research Fund 	16201120, in part by the 2020 Foshan HKUST Project FSPM02202009-2, and in part by the National Natural Science Foundation of China under grants 62073003 and 72131001. }}

\author{
Jianqi~Chen, \thanks{Jianqi Chen and Li Qiu are with the Department of Electronic and Computer
Engineering (ECE), The Hong Kong University of Science and Technology (HKUST), Clear
Water Bay, Kowloon, Hong Kong, China.
{\tt\small eejianqichen@ust.hk, eeqiu@ust.hk}}
Wei~Chen, \thanks{Wei Chen is with the Department of Mechanics and Engineering Science \& State Key Laboratory for Turbulence and Complex Systems, Peking University, Beijing 100871, China.
{\tt\small w.chen@pku.edu.cn}}
Chao~Chen, and 
Li Qiu,~\IEEEmembership{~Fellow IEEE}
\thanks{Chao~Chen was with the Department of ECE, HKUST, Hong Kong, China, and is now with the   Department of Electrical Engineering (ESAT), KU Leuven, 3001 Leuven, Belgium. {\tt\small cchenap@connect.ust.hk}}
}

 \IEEEoverridecommandlockouts
\begin{document}

\maketitle

\begin{abstract}

This study first introduces the frequency-wise phases of $n$-port linear time-invariant networks based on recently defined phases of complex matrices. 
Such a
phase characterization can be used to quantify the well-known notion of passivity
for networks. 
Further, a class of matrix operations induced by fairly common $n$-port network connections is examined.
The intrinsic phase properties of networks under such connections
are preserved. Concretely, a scalable phase-preserving criterion is proposed, which involves only
the phase properties of individual subnetworks, under the matrix operations featured by
connections. This criterion ensures that the phase range of the integrated network can be verified
effectively and that the scalability of the analyses can be maintained.
In addition, the inverse operations of the considered connections, that is, network subtractions with correspondences are examined.
With the known phase ranges of the integrated network and one of its subnetworks, the maximal allowable phase range of the remaining subnetwork can also be determined explicitly in a unified form for all types of subtractions. 
Finally, we extend the phase-preserving properties from the aforementioned connections to more general matrix operations defined using a certain indefinite inner product.  
\end{abstract}

\section{Introduction}
{L}{arge}-scale networks, including the transportation, communication, social,  power, and mechanical networks,  pervade various domains. The investigation of large-scale networks has been a longstanding and prominently focused research area over recent decades. 
In the study of such networks, there are often instances where one's primary focus is on characterizing their behavior through just an input-output perspective or solely at their boundary nodes, rather than delving into their internal structure. In such scenarios, a so-called \emph{port-based description} offers a useful network representation. More precisely, an $n$-port network is characterized by its port variables and engages in interactions with other $n$-port networks via \emph{port connections}.
It has proven to be suitable for elucidating diverse physical systems spanning electrical networks \cite{desoer1966basic,anderson2013network,laib2023decentralized}, mechanical networks \cite{smith2000performance,smith2002synthesis, chen2009missing}, and fluid and thermal systems \cite{shearer1967introduction,perelson1975network}. For example, an $n$-port electrical network with a potentially substantial internal scale can be described by the behavior of current-voltage pairs applied to its port variables. When an electrical network is linear and time-invariant (LTI), its port behavior can be characterized in the frequency domain using its associated impedance or admittance matrix (Z-matrix or Y-matrix, respectively); please refer to the following comprehensive texts: \cite{desoer1966basic, anderson2013network, youla2015theory}.
A proliferation of problems and active topics incorporating the operations of such $n$-port networks have been extensively investigated since the 1950s, such as network realizations, syntheses, robustness analyses, order reductions, and performance attainments \cite{shannon1949synthesis,biorci1961synthesis,carlin1964singular,youla2015theory,pozar2011microwave,odabasioglu1998prima,ortega2003power,garcia2007power,fiaz2013port,chaffey2021monotone,miranda2022dissipativity,laib2023decentralized,chaffey2023circuit}.

\emph{Scalability} is a desired property in the context of large-scale networks. It is important to develop compositional criteria that can ensure global properties and behaviours of the  networks based on the characteristics of individual components and their interactions in a well-scaled manner. The classical concepts of passivity, and more broadly, dissipativity, hold significant relevance in this regard \cite{desoer1966basic,anderson2013network,van2000l2}. They have been widely used in the study of large-scale networks, from classical topics like stability to more recent works on consensus, synchronization, and distributed control, etc. See for instance \cite{hill1980dissipative,arcak2007passivity,dorfler2014synchronization,ren2010distributed,
li2020distributed}. In the context of $n$-port networks, a well-known property is that when all subnetworks composed exclusively of passive components, such as resistors, inductors, and capacitors in a circuit network, are interconnected in arbitrary ways, the overall interconnected $n$-port network remains passive \cite{desoer1966basic}. 
Then a question emerges: If some subnetworks have a few non-passive elements such as dependent or independent sources, do the integrated networks preserve passivity under \emph{arbitrary interconnections}? Though this appears a straightforward question, it seems not to have been fully addressed. The study conducted in this paper will provide a comprehensive answer to this question in a broader context not restricted to passivity but rather extended to general phase-bounded networks.

he main focus of this work is on phases of $n$-port networks. Recently, there have been collective efforts in introducing a phase notion of MIMO LTI systems and studying the feedback system from a phasic perspective \cite{wang2020phases, chen2021phase, mao2022phases}. The phase of an $n$-port LTI network is defined
on its associated Z-matrix in the frequency domain. The phase concept enables one to quantify and extend passivity from a sector-wise perspective; that is, 
the phases of passive networks all lie within $[-\pi/2,\pi/2]$, which was only well understood in one-port passive networks in the past. Although newly introduced, the phase notion has been evidenced to play significant roles in developing scalable conditions on network stability and synchronization problems \cite{chen2020phase, chen2021phase, mao2022phases, Dan2023phase}. 
We mention that there have been numerous other efforts in studying phase information of MIMO systems. In an early attempt \cite{postlethwaite1981principal}, the principal phases were introduced from the unitary part of polar decomposition
of a system transfer function matrix. 
In \cite{reza1980concept} the energy phases for passive $n$-port networks composed of only passive components were defined in relation to the power factor of electrical networks.
Attempts have also been made to use the angles between the left and right singular vectors of the transfer function matrix to define phases, thereby ensuring a one-to-one correspondence between each magnitude and phase \cite{freudenberg1988frequency,chen1998multivariable}. 
The recently introduced phase in \cite{wang2020phases,chen2020phase,chen2021phase, mao2022phases} has advantages in quantifying and extending passivity like the role of phase in SISO systems and establishing a small phase theorem for feedback stability.

In this work, we are interested in using phase information of subnetworks or subsystems to ascertain phase properties of the overall networks under a broad range of network interconnections, so as to have a scale-free phase analysis in large-scale $n$-port networks.
In the literature of $n$-port networks, network interconnections from the relatively simple series and  parallel connections to complicated hybrid and cascade connections, and further, the more generalized 
hybrid-cascade connections, have all been subjects of study, with respect to both physical networks and associated matrix operator representations \cite{anderson1971shorted,anderson1975matrix,mitra1982shorted,mitra1975hybrid,anderson1986cascade,mitra1986parallel,mitra1983fundamental,pekarev1992shorts,antezana2006bilateral,zhao2019stabilization
,chen2020spectral}.
Notwithstanding, a nonnegligible gap appears to exist; that is, studies on exploring the phase properties of the $n$-port networks involving different connections are scarce owing to the lack of a suitable definition.
%
%
%
By virtue of the phase definition introduced in \cite{wang2020phases,chen2020phase}, under different matrix operations induced by the widespread network connections elaborated in \cite{anderson1975matrix,mitra1982shorted,mitra1975hybrid,anderson1986cascade,mitra1986parallel,mitra1983fundamental}, this study examines the phase range of the obtained $n$-port network after integrating several subnetworks subjected to phasic constraints.
Surprisingly,  a scalable phase-preserving criterion
is established successfully, indicating a hidden phasic relationship, i.e., the consensus in the attainable range of phases between a large-scale multiport network and its
interconnected components, respectively.

In contrast to connections, network subtractions as inverse operations are rarely mentioned in the literature. 
The subtractions can decompose an $n$-port network with a sophisticated higher-order internal structure into two or more $n$-port networks that admit relatively low-order internal components,
amenable to
certain rules. For example, series, parallel, hybrid, cascade, and hybrid-cascade subtractions were considered in \cite{mitra1982shorted,mitra1975hybrid,anderson1986cascade,mitra1986parallel,pekarev1992shorts, antezana2006bilateral}. Considering the phasic constraints of the primal $n$-port high-order network and that of the 
 subnetworks, the question that arises is can the allowable phase ranges of other subnetworks be determined? The answer is affirmative, and we obtain a closed-form phase range independent of the choice of subtraction.
 
Finally, we extend the network connections enumerated earlier into a unified framework involving matrix operations
defined on certain subspaces endowed with an indefinite inner product. The algebraic formulations used
imply a number of physical principles, such as Kirchhoff's voltage and current laws and the principle of Maxwell \cite{maxwell1873treatise}. 
Therefore,  all physically realizable network connections can be studied using this framework. Moreover, generalized scalable phase-preserving properties
 are addressed.

The remainder of this paper is organized as follows. In Section~\ref{sec: preliminaries}, preliminaries of the complex matrix phases are introduced. 
The $n$-port networks, including their representations in the frequency domain and the phase responses, are defined in Section~\ref{sec: Nport}.
The frequently encountered network connections are discussed in Section~\ref{sec:connections}, along with an in-depth investigation of the intrinsic phase-preserving properties of 
such connections.
In Section~\ref{sec: Subtraction}, network subtractions are studied using the associated phase analyses. 
The generalization of the preceding phase-preserving results is discussed in Section~\ref{sec:Extension}, which is accessible
 to the more general matrix operations defined over certain subspaces endowed with an indefinite inner product.
Further, certain illustrative examples are provided in Section~\ref{sec: example}.
Finally, Section~\ref{sec: conclusion} concludes this paper.

The notation used in this paper is generally standard and will be clarified as required. Partial results of this paper were previously presented in the conference \cite{chen2023port}.

\section{Preliminaries: Phases of A Complex Matrix}\label{sec: preliminaries}
\begin{figure}
\begin{center}
\includegraphics[width=4.0cm, height=3.5cm]{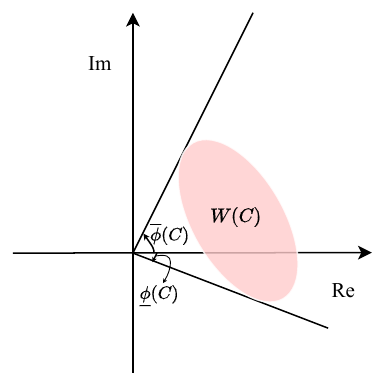}
\vspace{-0mm}
\caption{Geometric interpretation of $\overline{\phi}(C)$ and $\underline{\phi}(C)$.}
\label{Fig.phase}
\end{center}
\end{figure}

A nonzero complex scalar $c$, in polar form, can be represented as a combination of its magnitude (modulus) $\sigma^c$ and phase (argument) $\theta^c$, that is,
$c=\sigma^c e^{j\theta^c}$, where $\sigma^c>0$ and $\theta^c$ acquire values in half,
open $2\pi$-interval such as $[0,2\pi)$ or $(-\pi,\pi]$. 
From complex number $c$ to complex matrix $C\in \mathbb{C}^{n\times n}$, it is widely recognized that 
the corresponding magnitudes can be extended under the notion of singular values:  
$$\overline{\sigma}(C)=\sigma_1(C)\geq \sigma_2(C)\geq \cdots \geq \sigma_n(C)=\underline{\sigma}(C).$$ 
In comparison, defining the phases of a complex matrix $C$ appears to be an important question.
In this study, we introduced a recently discovered definition of matrix phases \cite{zhang2015matrix,wang2020phases}.
First, matrix $C\in \mathbb{C}^{n\times n}$ is said to be \textit{sectorial} if its numerical range \cite{roger1994topics}
$$W(C)=\left\lbrace x^{\ast}C x : x\in \mathbb{C}^n ~\mathrm{with}~\norm{x}=1\right\rbrace$$
does not contain the origin \cite{zhang2015matrix}.
It has been shown that the sectorial matrix $C$ is uniquely congruent to the diagonal unitary matrix
under a certain permutation. Therefore, such a matrix always allows factorization
\begin{equation}\label{eq:factorization}
C=T^{\ast}DT
\end{equation}
 with a nonsingular matrix $T$ and a diagonal unitary matrix $D$, which is referred to as sectorial decomposition. The phases of sectorial  matrix $C$ are defined as the arguments of the eigenvalues (i.e., diagonal elements) of $D$ and are denoted by 
$$\overline{\phi}(C) = \phi_1(C) \geq \phi_2(C)\geq\cdots\geq \phi_n(C) = \underline{\phi}(C).$$ 
A geometric interpretation of the phases is shown in Fig.~\ref{Fig.phase}. 
The phases $\overline{\phi}(C)$ and $\underline{\phi}(C)$ are the angles between the positive real axis and each of the two supporting rays of $W(C)$, whereas the remaining phases are located in between.  Clearly,  it holds that $\overline{\phi}(C)-\underline{\phi}(C)< \pi$.   The numerical range of the sectorial matrix is bounded by a convex cone in the complex plane.

\begin{figure}
\begin{center}
\hspace{1cm}
\includegraphics[width=7cm, height=3.5cm]{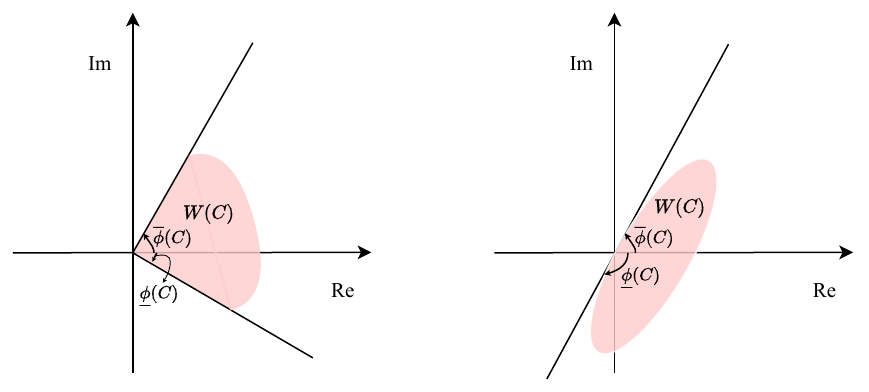}
\vspace{-0mm}
\caption{Geometric interpretation of semi-sectorial matrices.}
\label{Fig.semi-sectorial}
\end{center}
\end{figure}

For a non-sectorial matrix $C$,   its numerical range may be also bounded by a convex cone (see,    e.g. $\!\!\!$,   Fig.~\ref{Fig.semi-sectorial}).  
For example,  positive semi-definite matrices fall into this category.
Such a matrix $C$ is said to be semi-sectorial if the interior of its numerical range $W(C)$ does not contain zero.
A rank-$r$ semi-sectorial matrix $C$ admits the following generalized sectorial decomposition \cite{furtado2003spectral}:

\begin{equation}
C=T^{\ast}\begin{bmatrix}
0_{n-r}&0&0\\
0&D&0\\
0&0&E
\end{bmatrix}T,
\nonumber
\end{equation}
where
$$D=\mathrm{diag}\left\lbrace e^{j\theta_1},\ldots, e^{j\theta_m}\right\rbrace,$$
$$E=\mathrm{diag}\left\lbrace e^{j\theta_0}\begin{bmatrix}
1&2\\0&1
\end{bmatrix},\ldots, e^{j\theta_0}\begin{bmatrix}
1&2\\0&1
\end{bmatrix}
\right\rbrace\in \mathbb{C}^{(r-m)\times(r-m)},$$
and
$$\theta_0+\pi/2\geq \theta_1\geq \cdots \geq \theta_m\geq \theta_0+\pi/2.$$ 
Such a semi-sectorial matrix holds $r$ phases, including $\theta_1,\ldots,\theta_m$ and $(r-m)/2$ copies of $\theta_0\pm \pi/2$.
Further details on the comprehensive development of the matrix phase can be found in \cite{wang2020phases,qiu2022phases}.

\section{$N$-Port Network}\label{sec: Nport}
An $n$-port network serves as a fundamental framework for analyzing a wide range of  physical systems, encompassing electrical networks \cite{desoer1966basic,anderson2013network,laib2023decentralized}, mechanical networks \cite{smith2000performance,smith2002synthesis, chen2009missing}, and fluid and thermal systems
\cite{shearer1967introduction,perelson1975network}, by adopting an input-output perspective. In this section, we will primarily focus on 
$n$-port electrical networks as examples for illustration, reserving a discussion on 
$n$-port mechanical networks at the end of this section to provide supplementary insights.

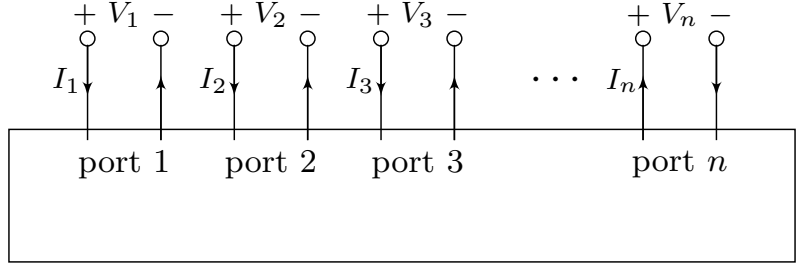
\begin{figure}[h]

\begin{center}

 \resizebox{0.6\columnwidth}{!}{
\begin{tikzpicture}[auto, node distance=3cm,>=latex']

\draw (-3.0,1.5) circle (2pt);
\draw (-2.3,1.5) circle (2pt);

\draw (-1.6,1.5) circle (2pt);
\draw (-0.9,1.5) circle (2pt);

\draw (-0.2,1.5) circle (2pt);
\draw (0.5,1.5) circle (2pt);

\draw (3.0,1.5) circle (2pt);
\draw (2.3,1.5) circle (2pt);

\node[draw, rectangle, minimum height=3em, minimum width=7.5cm ](Plant)
at (0,0) {};

\draw [-] (-3.0,1.44) -- (-3.0,0.53) {} ; 
\draw [->] (-3.0,1.44) -- (-3.0,0.95) {} ; 

\draw [-] (-2.3,1.44) -- (-2.3,0.53) {} ; 
\draw [->](-2.3,0.53) --(-2.3,1.15) {} ;

\draw [-] (-1.6,1.44) -- (-1.6,0.53) {} ; 
\draw [->] (-1.6,1.44) -- (-1.6,0.95) {} ; 

\draw [-] (-0.9,1.44) -- (-0.9,0.53) {} ; 
\draw [->](-0.9,0.53) --(-0.9,1.15) {} ;

\draw [-] (-0.2,1.44) -- (-0.2,0.53) {} ; 
\draw [->] (-0.2,1.44) -- (-0.2,0.95) {} ; 

\draw [-] (0.5,1.44) -- (0.5,0.53) {} ; 
\draw [->](0.5,0.53) --(0.5,1.15) {} ;

\draw [-] (3.0,1.44) -- (3.0,0.53) {} ; 
\draw [->] (3.0,1.44) -- (3.0,0.95) {} ; 

\draw [-] (2.3,1.44) -- (2.3,0.53) {} ; 
\draw [->](2.3,0.53) --(2.3,1.15) {} ; 

 \node  at (1.5,1.1) {$\cdots$};
 
 \node  at (-2.65,0.3) {{\footnotesize $\mathrm{port}~1$}};
 \node  at (-1.25,0.3) {{\footnotesize $\mathrm{port}~2$}};
 \node  at (0.15,0.3) {{\footnotesize $\mathrm{port}~3$}};
 \node  at (2.65,0.3) {{\footnotesize $\mathrm{port}~n$}};  
  
 \node  at (-3.2,1.1) {{\scriptsize $I_1$}};
 
 \node  at (-1.8,1.1) {{\scriptsize $I_2$}};
 
 \node  at (-0.4,1.1) {{\scriptsize $I_3$}};
 
 \node  at (2.1,1.1) {{\scriptsize $I_n$}};

  \node  at (-2.65,1.73) {{\scriptsize $+~V_1~-$}};
  
  \node  at (-1.25,1.73) {{\scriptsize $+~V_2~-$}};
  
  \node  at (0.15,1.73) {{\scriptsize $+~V_3~-$}};
  
  \node  at (2.65,1.73) {{\scriptsize $+~V_n~-$}};
\end{tikzpicture}
}
\end{center}
\caption{$n$-port network.}
\label{fig:n-port network}
\end{figure}

An $n$-port electrical network is a system comprising a connection of electrical components, such as resistors, inductors, capacitors, transformers, dependent or independent sources, and gyrators. An $n$-port network is associated with two $n$ external terminals paired to constitute ports. Each port satisfies the condition wherein the currents flowing into the two terminals are equal and opposite. These ports can be used to connect external circuit elements or ports of other networks. 
An illustration of an $n$-port network is shown in Fig.~\ref{fig:n-port network}.

\subsection{Matrix Representation}
In $n$-port network analysis, the external behavior of a network can be completely characterized using the voltages and currents at its ports. 
In this case, without specifying the internal structure of the network, a ``black box'' was applied to describe the network. 
A natural question is how it can be described properly using the relationship between the port voltage vector $V=[V_1~ V_2 \cdots V_n]^{\top}$ and port current vector $I=[I_1~ I_2 \cdots I_n]^{\top}$.
For simplicity, the networks under consideration are assumed to be linear, time-invariant, and composed of a finite number of lumped components. 
Subsequently, an $n$-port network may be described by its Z-matrix (impedance matrix) $Z$ relating $V$ and $I$ in the frequency domain \cite{pozar2011microwave}, that is,
\begin{equation}
V(s)=Z(s)I(s),
\nonumber
\end{equation}
where $I(s)$ is considered as an independent variable. 
For simplicity, we omit the variable $s$ when there is no confusion.

In addition to the Z-matrix,
there exist other matrices frequently used to describe the external behavior of $n$-port networks, such as the S-matrix, Y-matrix, H-matrix, T-matrix, and ABCD-matrix. 
The development of such representations is presented in \cite{desoer1966basic,pozar2011microwave,anderson2013network}.

\subsection{Phases of $N$-Port Network}
\begin{figure}[]
\begin{center}
\hspace{1cm}
\includegraphics[width=4.2cm, height=5.0cm]{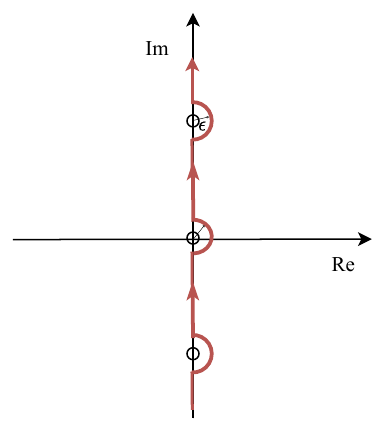}
\vspace{-0mm}
\caption{Indented  $j\omega$-axis:``$\circ$'' denotes the $j\omega$-axis poles of $Z$.}
\label{fig:Indented jomega-axis}
\end{center}
\end{figure}

Throughout this study, we considered $Z$ as the $n\times n$ real-rational transfer function matrix of an $n$-port network,
denoted by $Z\in \mathcal{R}^{n\times n}$, where
$ \mathcal{R}^{m\times n}$ represents the set of $m \times n$ real-rational transfers
function matrices.
Such an $n$-port network is assumed to be stable or semi-stable, that is, $Z$ has no pole in the open right-half plane. Moreover, $Z$ needs not to be proper for the inductor case to be appropriately addressed.
The frequency response of $Z$, denoted by $Z(j\omega)$, is defined along the indented  imaginary axis embedded in the 
half-circle detours with a sufficiently small radius $\epsilon>0$ around the unstable poles of $Z$ if they exist (Fig.~\ref{fig:Indented jomega-axis}).

The magnitude and phase of the $n$-port network can now be defined. 
The frequency-wise magnitudes of $Z$ are the singular values of its frequency response $Z(j\omega)$, bounded in the
interval $[\underline{\sigma}(Z(j\omega)),~\overline{\sigma}(Z(j\omega)]$. 

Analogously, it can develop frequency-wise phases of $Z$ by resorting to the preceding phase definition of complex matrices.
An $n$-port network $Z$ is considered as frequency-wise (semi-)sectorial if $Z(j\omega)$ is (semi-)sectorial for all $\omega\in\left[ -\infty,\infty\right] $.
For each fixed $\omega$, we let $\phi(Z(j\omega))$ be the vector of the phases of the (semi-)sectorial matrix $Z(j\omega)$. We can then define the set of phase-bounded (semi-)sectorial matrices
$$
\begin{aligned}
\mathcal{Z}[\alpha(\omega), \beta(\omega)&]:=\left\lbrace Z(j\omega)\!\in\! \mathbb{C}^{n\times n} \!:\! Z(j\omega)~\mathrm{is}~(\mathrm{semi}\textbf{-})\mathrm{sectorial}\right.\\ 
&\quad \left. \mathrm{and}~[\overline{\phi}(Z(j\omega)), \underline{\phi}(Z(j\omega))]\subset [\alpha(\omega), \beta(\omega)]\right\rbrace, 
\end{aligned}
$$
where $0\leq \beta(\omega)-\alpha(\omega)\leq \pi$.
Consider the next $\phi(Z(j\omega))$ as an element-wise function with $\omega\in\left[ -\infty,\infty\right] $, which is an odd function of frequency $\omega$. 
Hence,   we will focus on the positive frequency $\omega\in\left[ 0,\infty\right] $. 
We define the maximum and minimum phases of $Z(j\omega)$ over the interval $\left[ 0,\infty\right]$ as:
$$
\begin{aligned}
\overline{\phi}(Z)&=\sup\limits_{\omega\in [0,\infty]} \overline{\phi}(Z(j\omega))
\\
\underline{\phi}(Z)&=\inf\limits_{\omega\in [0,\infty]} \underline{\phi}(Z(j\omega)),
\end{aligned}
$$
and the  set of phase-bounded frequency-wise (semi-)sectorial networks:
$$
\begin{aligned}
\mathscr{Z}[\alpha, \beta&]:=\left\lbrace Z\!\in\!\mathcal{R}^{n\times n} \!:\! Z~\mathrm{is~frequency\textbf{-}wise}\right.\\ 
&\qquad \left. (\mathrm{semi}\textbf{-})\mathrm{sectorial~and}~[\overline{\phi}(Z), \underline{\phi}(Z)]\subset [\alpha, \beta]\right\rbrace
\end{aligned}
$$
for $0\leq \beta-\alpha\leq \pi.$
Note that in contrast to the frequency-wise $\mathcal{Z}[\alpha(\omega), \beta(\omega)]$, the set $\mathscr{Z}[\alpha, \beta]$ is independent of the frequency.

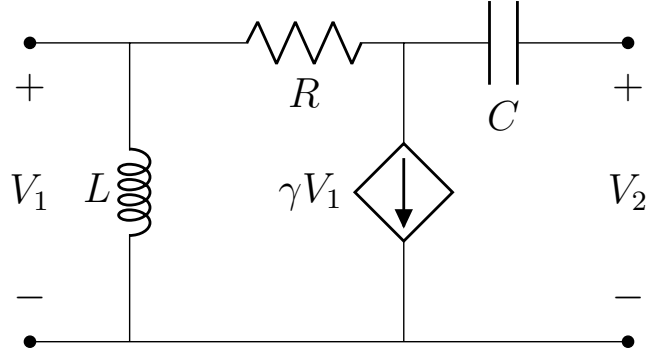
\begin{figure}
\begin{center}
 \resizebox{0.5\columnwidth}{!}{
\begin{circuitikz}[american voltages]
\draw
  (0,0) to [short, *-] (2,0)
  (3.5,0) to [short, -] (4,0)
  (5.5,0) to [short, -*] (6,0)
  (0,-3) to [short, *-*] (6,-3)
  (0,0) to [open, v^>=$V_1$] (0,-3) 
  (6,0) to [open, v^>=$V_2$] (6,-3)
  (2,0) to [R, l_=$R$] (3.5,0)
  (1,0) to [L, l_=$L$] (1,-3)
  (4,0) to [C, l_=$C$] (5.5,0)
  (3.75,0) to [american controlled current source, l_=$\gamma V_1$] (3.75,-3);
\end{circuitikz}
}
\end{center}
\caption{Two-port network.}
\label{fig:A 2-port network}
\end{figure}

\begin{figure}[h]
\begin{center}
\includegraphics[width=8.2cm, height=7.7cm]{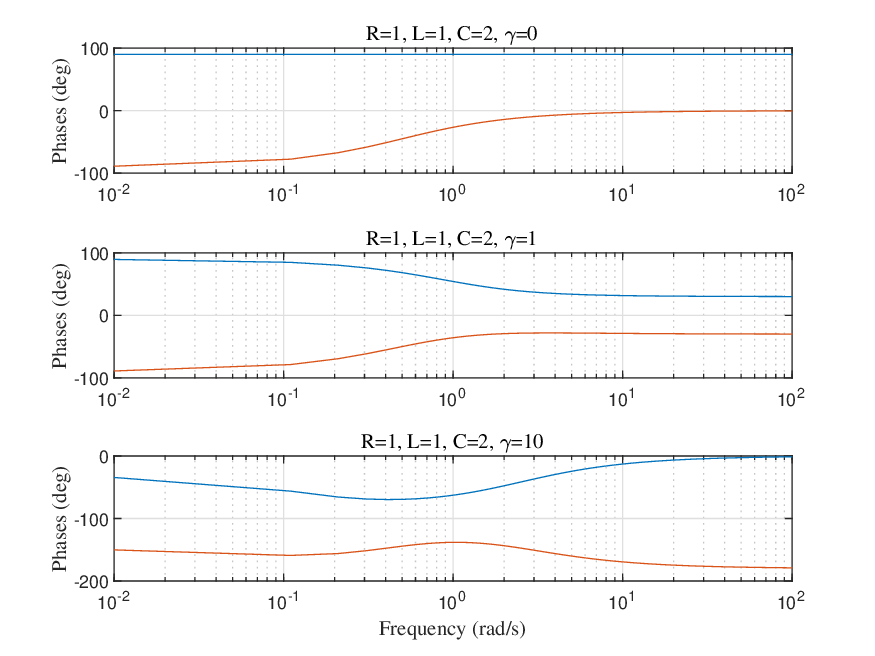}

\end{center}
\caption{Phase plots  of the frequency-wise (semi-)sectorial 2-port network $Z$ in \eqref{eq:2port}.}
\label{fig:generic phaseplot}
\end{figure}

Consider the 2-port network depicted in Fig.~\ref{fig:A 2-port network} as an illustrative example. The network involves
a resistor, inductor, capacitor, and current source dependent on $V_1$. The corresponding Z-matrix is expressed as:
\begin{equation}\label{eq:2port}
Z(s)=\dfrac{1}{1+\gamma Ls}
\begin{bmatrix}
Ls&Ls
\\
(1-\gamma R)Ls&\frac{1+\gamma Ls}{C s}+(Ls+R)
\end{bmatrix}.
\end{equation}
We take different parameters $\gamma$.   The corresponding networks satisfy
\begin{equation}
Z\in \left\lbrace 
\begin{aligned}
&\mathscr{Z}[-\pi/2, \pi/2],\qquad \gamma=0,
\\
&\mathscr{Z}[-\pi/2, \pi/2],\qquad \gamma=1,
\\
&\mathscr{Z}[-\pi, 0],\qquad\qquad \gamma=10.
\end{aligned}
\right. 
\nonumber
\end{equation}
The phase plots are shown in Fig.~\ref{fig:generic phaseplot}.
In addition,  for $\gamma=1$, it holds that
\begin{equation}
Z(j\omega)\in \left\lbrace 
\begin{aligned}
&\mathcal{Z}[-\pi/2, \pi/2],\qquad \forall \omega\in [0,10],
\\
&\mathcal{Z}[-\pi/4, \pi/4],\qquad \forall \omega\in (10,\infty].
\end{aligned}
\right. 
\nonumber
\end{equation}

\subsection{Passive Networks}\label{subsec:passive network}
An $n$-port resistive network is a circuit comprising only positive resistors, which is an important
class of frequently encountered electrical networks. The Z-matrix of any resistive network, denoted by $Z^r$, is a real positive semidefinite matrix \cite{biorci1961synthesis}.
Therefore, all phases of $Z^r$ are zero.

The $n$-port passive networks, including $n$-port resistive networks, can be synthesized by assembling passive components, such as positive resistors, capacitors, inductors, and other possible passive sub-networks.
For a passive network, the associated Z-matrix $Z$ exhibits the positive real property \cite{anderson2013network}; that is, 
$\mathrm{Re}\left( \left\langle Z(j\omega)x,x \right\rangle\right) =\mathrm{Re} \left( x^{\ast}Z(j\omega)x\right) \geq 0,$ for all $\omega\in [0,\infty]$ and $x\in \mathbb{C}^n.$ Here, $\mathrm{Re}(\cdot)$ represents the real part.
The phases of $Z$ always satisfy the following condition:

\begin{equation}\label{eq:passive 2}
Z\in \mathscr{Z}[-\pi/2, \pi/2].
\end{equation}

A concept that is closely related to passivity is losslessness. 
A lossless network is always passive and cannot adsorb any power injected into a network \cite{anderson2013network,georgiou2019principles}.
Such a network contains only lossless components, excluding resistors. Further, all elements of its Z-matrix are purely imaginary. Therefore, the phases are confined to the $-\pi/2$ and $\pi/2$.

\subsection{Mechanical Networks}
The $n$-port network representation can also be employed to characterize mechanical networks, wherein each port exerts an equal and opposite force $F$ and undergoes a relative velocity~$v$.
There exist several standard analogies between mechanical and electrical networks. 
Specifically,
the basic modeling elements of these two networks have correspondences: spring and inductor, damper and resistor, and 
mass and capacitor. In \cite{smith2000performance,smith2002synthesis, chen2009missing,smith2020inerter}, Smith et al. first proposed the ``inerter'' with two terminals replacing the one-terminal mass 
as a counterpart of the capacitor. Therefore, reinforced correspondence implies that any $n$-port passive electrical network can be 
mapped onto a realizable $n$-port passive mechanical network \cite{chen2009missing,georgiou2019principles}.

\section{Connections of Networks}\label{sec:connections}
In the remainder of this paper, we omit the variable $j\omega$ if there is no confusion, and thus, all the following relationships and results hold for all $\omega$.

First, we consider two one-port networks $A$ and $B$ with frequency responses $z^a=\sigma^a e^{j\theta^a}$ and $z^b=\sigma^b e^{j\theta^b}$, respectively.
If two networks are connected in series, the frequency response of the 
 a one-port network $C$ is expressed as
\begin{equation}\label{eq:c=a+b}
z^c=z^a+z^b. 
\end{equation} 
 A natural step forward is the exploration of the magnitude (phase) response relationship between $\sigma^c \left( \theta^c \right) $ and
 $\sigma^a ,~\sigma^b \left(\theta^a ,~\theta^b \right).$  Based on the triangle inequality, it follows 
  \begin{equation}
  \sigma^c\leq \sigma^a+ \sigma^b.
  \end{equation}
 Regarding the phase responses, a tedious proof yields the following result:
 \begin{proposition}\label{prop:1}
 If $|\theta^a-\theta^b|\leq \pi$, then:
 \begin{equation}\label{eq:phase relationship SISO connection}
\begin{aligned}
&\theta^c \in \left[\min\left\lbrace \theta^a,\theta^b \right\rbrace, \max\left\lbrace \theta^a,\theta^b \right\rbrace \right].
\end{aligned} 
 \end{equation}
 \end{proposition}

This study was more focused on a phase relationship similar to \eqref{eq:phase relationship SISO connection}, which motivates our phase research on $n$-port networks.
Exploring methods of extending it
to $n$-port networks under various types of connections is the primary purpose of this section.
First, we address commonly encountered forms of connections among $n$-port networks and their associated matrix operations in terms of the corresponding Z-matrices, which were also introduced in \cite{anderson1975matrix,mitra1975hybrid,duffin1978almost,anderson1986cascade,mitra1982shorted}.
All the connections and then results obtained in the following sections remain applicable to all types of
$n$-port networks. For the sake of illustration, we have opted to utilize the symbols (current $I$ and voltage $V$) of electrical networks. We begin with the simplest shorted connection.

\textbf{(i) Shorted connection:}
For an $n$-port network $A$, as illustrated in Fig.~\ref{fig:n-port network}, associated with the Z-matrix $Z^a$, its ports can be separated 
into two groups: the first $r$ ports, and the remaining $(n\!-\!r)$ ports. Accordingly, we can partition $Z^a$ as 
\begin{equation}\label{eq:Partition Z}
Z^a=\begin{bmatrix}
Z_{11}^a & Z_{12}^a\\
Z_{21}^a & Z_{22}^a
\end{bmatrix}\in \mathcal{R}^{\left( r+(n-r)\right) \times \left( r+(n-r)\right) }.
\end{equation}

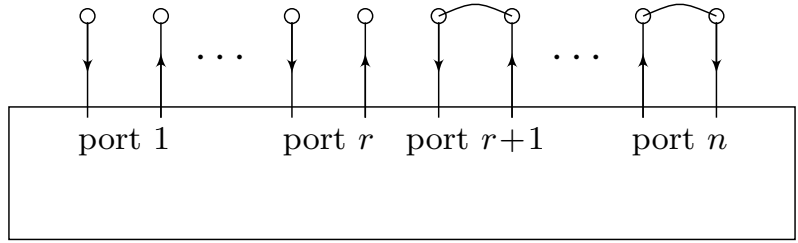
\begin{figure}[h]
\begin{center}
 \resizebox{0.6\columnwidth}{!}{
\begin{tikzpicture}[auto, node distance=3cm,>=latex']

\draw (-3.0,1.5) circle (2pt);
\draw (-2.3,1.5) circle (2pt);

\draw (-1.05,1.5) circle (2pt);
\draw (-0.35,1.5) circle (2pt);

\draw (1.05,1.5) circle (2pt);
\draw (0.35,1.5) circle (2pt);

\draw (3.0,1.5) circle (2pt);
\draw (2.3,1.5) circle (2pt);

\node[draw, rectangle, minimum height=3em, minimum width=7.5cm ](Plant)
at (0,0) {};

\draw [-] (-3.0,1.44) -- (-3.0,0.53) {} ; 
\draw [->] (-3.0,1.44) -- (-3.0,0.95) {} ; 

\draw [-] (-2.3,1.44) -- (-2.3,0.53) {} ; 
\draw [->](-2.3,0.53) --(-2.3,1.15) {} ;

\draw [-] (-1.05,1.44) -- (-1.05,0.53) {} ; 
\draw [->] (-1.05,1.44) -- (-1.05,0.95) {} ; 

\draw [-] (-0.35,1.44) -- (-0.35,0.53) {} ; 
\draw [->](-0.35,0.53) --(-0.35,1.15) {} ;

\draw [-] (0.35,1.44) -- (0.35,0.53) {} ; 
\draw [->] (0.35,1.44) -- (0.35,0.95) {} ; 

\draw [-] (1.05,1.44) -- (1.05,0.53) {} ; 
\draw [->](1.05,0.53) --(1.05,1.15) {} ;

\draw [-] (3.0,1.44) -- (3.0,0.53) {} ; 
\draw [->] (3.0,1.44) -- (3.0,0.95) {} ; 

\draw [-] (2.3,1.44) -- (2.3,0.53) {} ; 
\draw [->](2.3,0.53) --(2.3,1.15) {} ;

 \node  at (1.7,1.1) {$\cdots$};
 
 \node  at (-1.7,1.1) {$\cdots$};
 
 \node  at (-2.65,0.3) {{\footnotesize $\mathrm{port}~1$}};
 \node  at (-0.7,0.3) {{\footnotesize $\mathrm{port}~r$}};
 \node  at (0.7,0.3) {{\footnotesize $\mathrm{port}~r\!+\!1$}};
 \node  at (2.65,0.3) {{\footnotesize $\mathrm{port}~n$}};

\draw[-] (2.3,1.5) .. controls (2.65,1.65) .. (3,1.5);
\draw[-] (0.35,1.5) .. controls (0.7,1.65) .. (1.05,1.5);
\end{tikzpicture}
}
\end{center}
\caption{Shorted connection}
\label{fig:shorted connection}
\end{figure}

If the last $(n\!-\!r)$ ports of $A$ are shorted, as manifested  in Fig.~\ref{fig:shorted connection}, denoted by the shorted $r$-port network $C$, the following is obtained
\begin{equation}
\begin{aligned}
& V_i^c=V_i^a,~~~I_i^c=I_i^a,~~&i=1,2,\ldots,r
\\
& V_i^a=0,~~&i=r+1,r+2,\ldots,n.
\end{aligned}
\end{equation}

It is well-known from \cite{anderson1971shorted,anderson1975shorted,mitra1982shorted} that $Z^c$ exists if and only if the following conditions hold:
\begin{equation}\label{eq:well-posedness}
\mathcal{R}(Z^a_{21})\subseteq \mathcal{R}(Z^a_{22})~~\mathrm{and}~~\mathcal{N}(Z^{a}_{12})\supseteq \mathcal{N}(Z^{a}_{22}).
\end{equation}
Here $\mathcal{R}(\cdot)$ and $\mathcal{N}(\cdot)$ denote the column space and the kernel of a given matrix, respectively.
In such a case, 
\begin{equation}\label{eq:Schur Complement}
Z^c=Z^a/_{22}=Z^a_{11}-Z^a_{12}Z^{a^{\dagger}}_{22}Z^a_{21},
\end{equation}
where  $\dagger$ represents the Moore-Penrose inverse of $Z^{a}_{22}$ and $Z^a/_{22}$ is the Schur complement of $Z^a$.
The conditions \eqref{eq:well-posedness} have also been used to name the \textit{well-definedness} of the Schur complement in \cite{burns1974generalized} to
ensure that $Z^a/_{22}$ is independent of the choice of the generalized inverse.

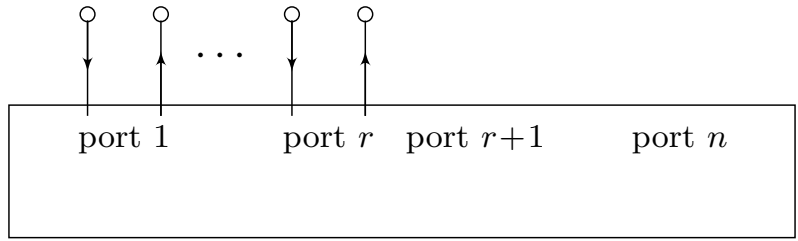
\begin{figure}[h]
\begin{center}
 \resizebox{0.6\columnwidth}{!}{
\begin{tikzpicture}[auto, node distance=3cm,>=latex']

\draw (-3.0,1.5) circle (2pt);
\draw (-2.3,1.5) circle (2pt);

\draw (-1.05,1.5) circle (2pt);
\draw (-0.35,1.5) circle (2pt);

\node[draw, rectangle, minimum height=3em, minimum width=7.5cm ](Plant)
at (0,0) {};

\draw [-] (-3.0,1.44) -- (-3.0,0.53) {} ; 
\draw [->] (-3.0,1.44) -- (-3.0,0.95) {} ; 

\draw [-] (-2.3,1.44) -- (-2.3,0.53) {} ; 
\draw [->](-2.3,0.53) --(-2.3,1.15) {} ;

\draw [-] (-1.05,1.44) -- (-1.05,0.53) {} ; 
\draw [->] (-1.05,1.44) -- (-1.05,0.95) {} ; 

\draw [-] (-0.35,1.44) -- (-0.35,0.53) {} ; 
\draw [->](-0.35,0.53) --(-0.35,1.15) {} ;

 \node  at (-1.7,1.1) {$\cdots$};
 
 \node  at (-2.65,0.3) {{\footnotesize $\mathrm{port}~1$}};
 \node  at (-0.7,0.3) {{\footnotesize $\mathrm{port}~r$}};
 \node  at (0.7,0.3) {{\footnotesize $\mathrm{port}~r\!+\!1$}};
 \node  at (2.65,0.3) {{\footnotesize $\mathrm{port}~n$}};

\end{tikzpicture}
}
\end{center}
\caption{Open connection}
\label{fig:open connection}
\end{figure}

\textbf{(ii) Open connection:}
If the last $n-r$ ports of $n$-port network A are left open, as shown in Fig.~\ref{fig:open connection},
that is,
\begin{equation}
\begin{aligned}
& V_i^c=V_i^a,~~~I_i^c=I_i^a,~~&i=1,2,\ldots,r
\\
&~ I_i^a=0,~~~&i=r+1,r+2,\ldots,n,
\end{aligned}
\end{equation}
 another $r$-port network $C$ is obtained whose $Z$-matrix is
\begin{equation}\label{eq:Zc open connection}
Z^c=Z^a_{11}.
\end{equation}

\begin{figure}[h]
\begin{center}
 \resizebox{0.60\columnwidth}{!}{
\begin{tikzpicture}[auto, node distance=3cm,>=latex']

\draw (-3.2,1.45) circle (2pt);
\draw (-2.1,1.0) circle (2pt);

\draw (-1.8,1.45) circle (2pt);
\draw (-0.7,1.0) circle (2pt);

\draw (-0.4,1.45) circle (2pt);
\draw (0.7,1.0) circle (2pt);

\draw (2.1,1.45) circle (2pt);
\draw (3.2,1.0) circle (2pt);

\node[draw, rectangle, minimum height=3em, minimum width=7.5cm ](Plant)
at (0,0) {};
\node  at (0,-0.15) {$\mathrm{Network~B}$};
 
\node[draw, rectangle, minimum height=3em, minimum width=7.5cm ](Plant)
at (0,2.5) {};
\node  at (0,2.65) {$\mathrm{Network~A}$};

\draw [-] (-3.2,1.45)--(-3.0,1.6){}; 
\draw [->](-3.0,1.6)--(-3.0,2.0){}; 

\draw [-] (-2.3,0.53)--(-2.3,0.75) -- (-2.1,1.0) {}; 
 
\draw [-] (-2.3,2)--(-2.3,1.6) -- (-3,0.85) {}; 
\draw [->](-3,0.85) --(-3.0,0.53){};

\draw [-] (-1.8,1.45)--(-1.6,1.6){}; 
\draw [->](-1.6,1.6)--(-1.6,2.0){}; 

\draw [-] (-0.9,0.53)--(-0.9,0.75) -- (-0.7,1.0) {}; 
 
\draw [-] (-0.9,2)--(-0.9,1.6) -- (-1.6,0.85) {}; 
\draw [->](-1.6,0.85) --(-1.6,0.53){};

\draw [-] (-0.4,1.45)--(-0.2,1.6){}; 
\draw [->](-0.2,1.6)--(-0.2,2.0){}; 

\draw [-] (0.5,0.53)--(0.5,0.75) -- (0.7,1.0) {}; 
 
\draw [-] (0.5,2)--(0.5,1.6) -- (-0.2,0.85) {}; 
\draw [->](-0.2,0.85) --(-0.2,0.53){};

\draw [-] (2.1,1.45)--(2.3,1.6){}; 
\draw [->](2.3,1.6)--(2.3,2.0){}; 

\draw [-] (3,0.53)--(3,0.75) -- (3.2,1.0) {}; 
 
\draw [-] (3,2)--(3,1.6) -- (2.3,0.85) {}; 
\draw [->](2.3,0.85) --(2.3,0.53){};

 \node  at (1.5,1.25) {$\cdots$};
 
 \node  at (-2.65,0.3) {{\footnotesize $\mathrm{port}~1$}};
 \node  at (-1.25,0.3) {{\footnotesize $\mathrm{port}~2$}};
 \node  at (0.15,0.3) {{\footnotesize $\mathrm{port}~3$}};
 \node  at (2.65,0.3) {{\footnotesize $\mathrm{port}~n$}}; 
 
  \node  at (-2.65,2.2) {{\footnotesize $\mathrm{port}~1$}};
 \node  at (-1.25,2.2) {{\footnotesize $\mathrm{port}~2$}};
 \node  at (0.15,2.2) {{\footnotesize $\mathrm{port}~3$}};
 \node  at (2.65,2.2) {{\footnotesize $\mathrm{port}~n$}}; 
 
\end{tikzpicture}
}
\end{center}
\caption{Series connection}
\label{fig:series connection}
\end{figure}
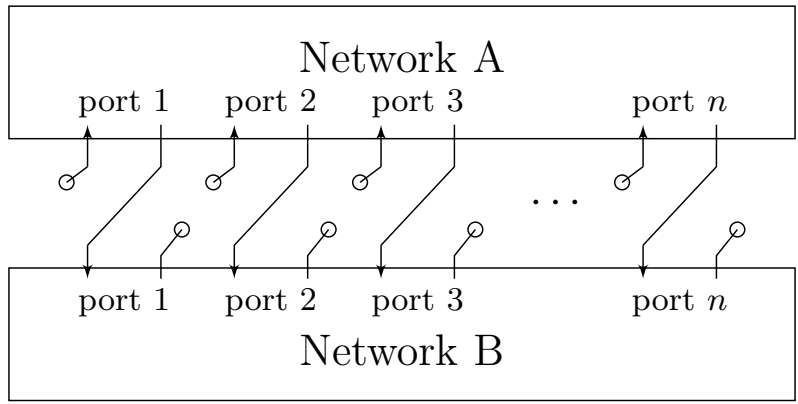

\textbf{(iii) Series connection:}
Consider two $n$-port networks $A$ and $B$. Let $C$ be an $n$-port network obtained via a series connection of $A$ and $B$ as
illustrated in Fig.~\ref{fig:series connection}. In this case, 
\begin{equation}
V_i^c=V_i^a+V_i^b,~~~I_i^c=I_i^a=I_i^b,~~i=1,2,\ldots,n,
\end{equation} 
and
\begin{equation}\label{eq:Zc Series connection}
Z^c=Z^a+Z^b.
\end{equation}

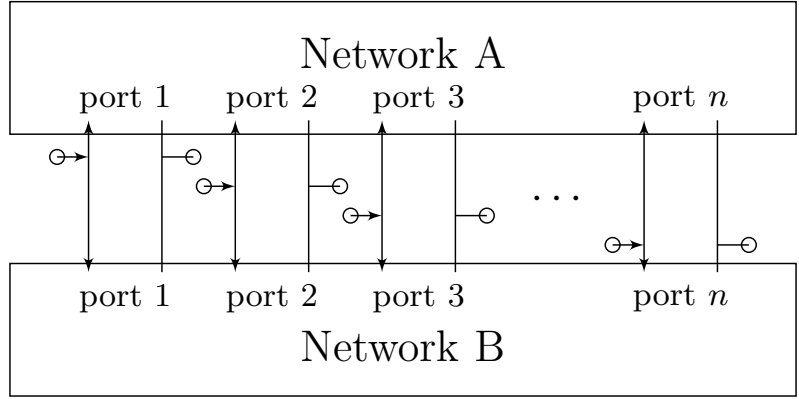
\begin{figure}[h]
\begin{center}
 \resizebox{0.60\columnwidth}{!}{
\begin{tikzpicture}[auto, node distance=3cm,>=latex']

\draw (-3.3,1.65) circle (2pt);
\draw (-2.0,1.65) circle (2pt);

\draw (-1.9,1.37) circle (2pt);
\draw (-0.6,1.37) circle (2pt);

\draw (-0.5,1.09) circle (2pt);
\draw (0.8,1.09) circle (2pt);

\draw (2.0,0.81) circle (2pt);
\draw (3.3,0.81) circle (2pt);

\node[draw, rectangle, minimum height=3em, minimum width=7.5cm ](Plant)
at (0,0) {};
\node  at (0,-0.15) {$\mathrm{Network~B}$};
 
\node[draw, rectangle, minimum height=3em, minimum width=7.5cm ](Plant)
at (0,2.5) {};
\node  at (0,2.65) {$\mathrm{Network~A}$};

\draw [<->] (-3,0.55)--(-3,2){}; 
\draw [-] (-2.3,0.55)--(-2.3,2){}; 

\draw [<->] (-1.6,0.55)--(-1.6,2){}; 
\draw [-] (-0.9,0.55)--(-0.9,2){}; 

\draw [<->] (-0.2,0.55)--(-0.2,2){}; 
\draw [-] (0.5,0.55)--(0.5,2){}; 

\draw [<->] (2.3,0.55)--(2.3,2){}; 
\draw [-] (3,0.55)--(3,2){}; 
\draw [->](-3.3,1.65)--(-3,1.65){};
\draw [-](-2.3,1.65)--(-2.0,1.65){};

\draw [->](-1.9,1.37)--(-1.6,1.37){};
\draw [-](-0.9,1.37)--(-0.6,1.37){};

\draw [->](-0.5,1.09)--(-0.2,1.09){};
\draw [-](0.5,1.09)--(0.8,1.09){};

\draw [->](2.0,0.81)--(2.3,0.81){};
\draw [-](3,0.81)--(3.3,0.81){};

 \node  at (1.5,1.25) {$\cdots$};
 
 \node  at (-2.65,0.3) {{\footnotesize $\mathrm{port}~1$}};
 \node  at (-1.25,0.3) {{\footnotesize $\mathrm{port}~2$}};
 \node  at (0.15,0.3) {{\footnotesize $\mathrm{port}~3$}};
 \node  at (2.65,0.3) {{\footnotesize $\mathrm{port}~n$}}; 
 
  \node  at (-2.65,2.2) {{\footnotesize $\mathrm{port}~1$}};
 \node  at (-1.25,2.2) {{\footnotesize $\mathrm{port}~2$}};
 \node  at (0.15,2.2) {{\footnotesize $\mathrm{port}~3$}};
 \node  at (2.65,2.2) {{\footnotesize $\mathrm{port}~n$}}; 
 
\end{tikzpicture}
}
\end{center}
\caption{Parallel connection}
\label{fig:parallel connection}
\end{figure}

\textbf{(iv) Parallel connection:}
An $n$-port network $C$ can be constructed by connecting $A$ and $B$ in parallel (Fig.~\ref{fig:parallel connection}). 
Formally, we obtain
\begin{equation}
V_i^c=V_i^a=V_i^b,~~~I_i^c=I_i^a+I_i^b,~~i=1,2,\ldots,n,
\end{equation} 
and
\begin{equation}
Z^c=Z^a \vcentcolon Z^b=Z^{a}\left(Z^{a}+Z^{b}\right)^\dagger Z^{b}.
\end{equation}

\begin{figure}[h]
\begin{center}
 \resizebox{0.60\columnwidth}{!}{
\begin{tikzpicture}[auto, node distance=3cm,>=latex']

\draw (-3.2,1.45) circle (2pt);
\draw (-2.1,1.) circle (2pt);

\draw (-1.34,1.45) circle (2pt);
\draw (-0.05,1.) circle (2pt);

\draw (0.05,1.45) circle (2pt);
\draw (1.35,1.45) circle (2pt);

\draw (2.0,1.) circle (2pt);
\draw (3.3,1.) circle (2pt);
\node[draw, rectangle, minimum height=3em, minimum width=7.5cm ](Plant)
at (0,0) {};
\node  at (0,-0.15) {$\mathrm{Network~B}$};
 
\node[draw, rectangle, minimum height=3em, minimum width=7.5cm ](Plant)
at (0,2.5) {};
\node  at (0,2.65) {$\mathrm{Network~A}$};

\draw [-] (-3.2,1.45)--(-3.0,1.6){}; 
\draw [->](-3.0,1.6)--(-3.0,2.0){}; 

\draw [-] (-2.3,0.53)--(-2.3,0.75) -- (-2.1,1.0) {}; 
 
\draw [-] (-2.3,2)--(-2.3,1.6) -- (-3,0.85) {}; 
\draw [->](-3,0.85) --(-3.0,0.53){}; 

\draw [-] (-1.34,1.45)--(-1.14,1.6){}; 
\draw [->](-1.14,1.6)--(-1.14,2.0){}; 

\draw [-] (-0.25,0.53)--(-0.25,0.76) -- (-0.05,1.) {}; 
 
\draw [-] (-0.25,2)--(-0.25,1.6) -- (-1.14,0.85) {}; 
\draw [->](-1.14,0.85) --(-1.14,0.53){};

\draw [<->] (0.35,0.55) -- (0.35,2) {} ; 
\draw [-] (1.05,0.55) -- (1.05,2) {} ;

\draw [<->] (3.0,0.55) -- (3.0,2) {} ; 
\draw [-] (2.3,0.55) -- (2.3,2) {} ;


\draw [->] (0.05,1.45) -- (0.35,1.45) {} ; 
\draw [-] (1.05,1.45) -- (1.35,1.45) {} ;

\draw [->] (2,1) -- (2.3,1) {} ; 
\draw [-] (3,1) -- (3.3,1) {} ; 
 \node  at (1.7,1.1) {$\cdots$};
 
 \node  at (-1.7,1.1) {$\cdots$};
 \node  at (-2.65,0.3) {{\footnotesize $\mathrm{port}~1$}};
 \node  at (-0.7,0.3) {{\footnotesize $\mathrm{port}~r$}};
 \node  at (0.7,0.3) {{\footnotesize $\mathrm{port}~r\!+\!1$}};
 \node  at (2.65,0.3) {{\footnotesize $\mathrm{port}~n$}};  
 \node  at (-2.65,2.2) {{\footnotesize $\mathrm{port}~1$}};
 \node  at (-0.7,2.2) {{\footnotesize $\mathrm{port}~r$}};
 \node  at (0.7,2.2) {{\footnotesize $\mathrm{port}~r\!+\!1$}};
 \node  at (2.65,2.2) {{\footnotesize $\mathrm{port}~n$}};

\end{tikzpicture}
}
\end{center}
\caption{Hybrid connection}
\label{fig:hybrid connection}
\end{figure}
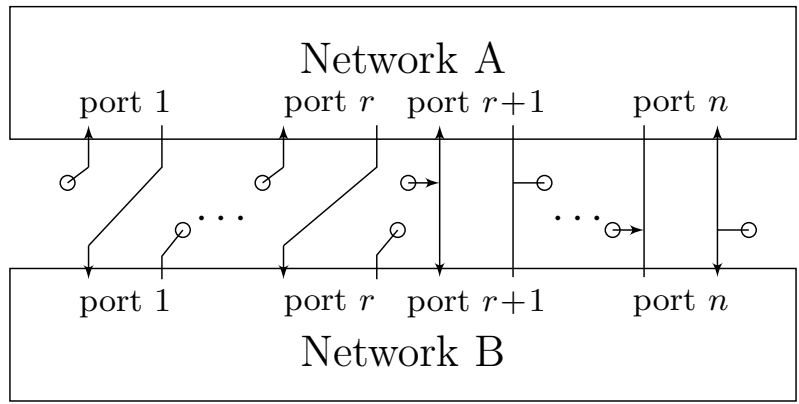

\textbf{(v) Hybrid connection:}
The hybrid connection is a mixture of series and parallel connections and contains each as a special case.  Figure \ref{fig:hybrid connection} illustrates the hybrid connection
between $n$-port networks $A$ and $B$.  For this connection,  we have 
\begin{equation}
\begin{aligned}
&V_i^c=V_i^a+V_i^b,~~~I_i^c=I_i^a=I_i^b,~~i=1,2,\ldots,r,
\\
&V_i^c=V_i^a=V_i^b,~~~I_i^c=I_i^a+I_i^b,~~i=r+1,r+2,\ldots,n.
\end{aligned}
\end{equation}
Thus, the hybrid connection also suggests a matrix operation,
denoted by
\begin{equation}\label{eq:Hybrid Zc1}
Z^c=Z^a \ast Z^b,
\end{equation}
where the explicit expression of $Z^c$ in \eqref{eq:Hybrid Zc1} is elaborated in Appendix A for brevity.

\begin{figure}[h]
\begin{center}
 \resizebox{0.6\columnwidth}{!}{
\begin{tikzpicture}[auto, node distance=3cm,>=latex']

\draw (-4,1.05) circle (2pt);
\draw (-4,0.65) circle (2pt);

\draw (-4,-1.05) circle (2pt);
\draw (-4,-0.65) circle (2pt);

\node  at (-4,0) {$\vdots$};

\draw (4,1.05) circle (2pt);
\draw (4,0.65) circle (2pt);

\draw (4,-1.05) circle (2pt);
\draw (4,-0.65) circle (2pt);

\node  at (4,0) {$\vdots$};

\node  at (0,0) {$\vdots$};
\node[draw, rectangle, minimum height=2.5cm, minimum width=2.7cm ](Plant)
at (-2.05,0) {$\mathrm{Network~A}$};

\node[draw, rectangle, minimum height=2.5cm, minimum width=2.7cm ](Plant)
at (2.05,0) {$\mathrm{Network~B}$};

\draw [->] (-4,1.05) -- (-3.4,1.05) {} ; 
\draw [-] (-4,0.65) -- (-3.4,0.65) {} ;

 \draw [-] (-4,-1.05) -- (-3.4,-1.05) {} ; 
\draw [->] (-4,-0.65) -- (-3.4,-0.65) {} ;

\draw [<-] (3.4,1.05) -- (4,1.05) {} ; 
\draw [-] (3.4,0.65) -- (4,0.65) {} ;

 \draw [-] (3.4,-1.05) -- (4,-1.05) {} ; 
\draw [<-] (3.4,-0.65) -- (4,-0.65) {} ;

\draw [-] (-0.7,-1.05) -- (0.7,-1.05) {} ; 
\draw [<->] (-0.7,-0.65) -- (0.7,-0.65) {} ;

\draw [<->] (-0.7,1.05) -- (0.7,1.05) {} ; 
\draw [-] (-0.7,0.65) -- (0.7,0.65) {} ;  

 \node  at (-2.9,0.85) {{\footnotesize $\mathrm{port}~1$}};
 \node  at (-2.9,-0.85) {{\footnotesize $\mathrm{port}~r$}};
 \node  at (-1.35,0.85) {{\footnotesize $\mathrm{port}~r\!+\!1$}};
 \node  at (-1.35,-0.85) {{\footnotesize $\mathrm{port}~r\!+\!s$}};

 \node  at (2.75,0.85) {{\footnotesize $\mathrm{port}~s\!+\!1$}};
 \node  at (2.75,-0.85) {{\footnotesize $\mathrm{port}~s\!+\!t$}};
 \node  at (1.2,0.85) {{\footnotesize $\mathrm{port}~1$}};
 \node  at (1.2,-0.85) {{\footnotesize $\mathrm{port}~s$}};
\end{tikzpicture}
}
\end{center}
\caption{Cascade connection}
\label{fig:cascade connection}
\end{figure}
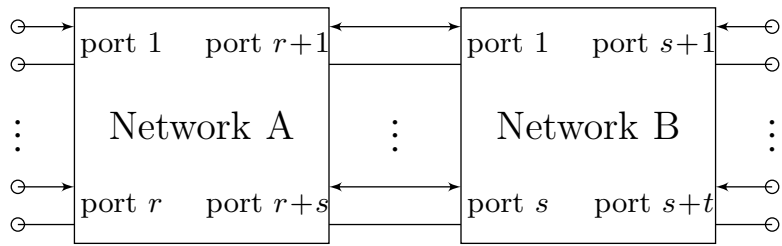

\textbf{(vi) Cascade connection:}
Considering the configuration shown in Fig.~\ref{fig:cascade connection}, where $A$ and $B$ are $(r+s)$-port and $(s+t)$-port networks, respectively,  and the resulted  $C$ is an $(r+t)$-port network.
The equations governing the port currents and voltages as shown in Fig.~\ref{fig:cascade connection} are 
\begin{equation}
\begin{aligned}
&V_i^c=V_i^a,          &I_i^c=I_i^a,~~~~~~&i=1,2,\ldots,r,
\\
&V_{r+i}^a=V_i^b,      &I_{r+i}^a=-I_i^b, ~~&i=1,2,\ldots,s,
\\
&V_{r+i}^c=V_{s+i}^b,  &I_{r+i}^c=I_{s+i}^b, ~~&i=r+1,r+2,\ldots,t.
\end{aligned}
\end{equation}
The matrix operation is expressed as:
\begin{equation}\label{eq:Zc Cascade}
Z^c=Z^a \circ Z^b.
\end{equation}
Appendix A presents the explicit expression of $Z^c$ in \eqref{eq:Zc Cascade}.

\begin{figure}[h]
\begin{center}
 \resizebox{0.6\columnwidth}{!}{
\begin{tikzpicture}[auto, node distance=3cm,>=latex']

\draw (-4.5,1.05) circle (2pt);
\draw (-4.5,0.65) circle (2pt);

\draw (-4.5,-1.05) circle (2pt);
\draw (-4.5,-0.65) circle (2pt);

\node  at (-4.5,0) {$\vdots$};

\draw (4.0,1.05) circle (2pt);
\draw (4.0,0.65) circle (2pt);

\draw (4.0,-1.05) circle (2pt);
\draw (4.0,-0.65) circle (2pt);

\node  at (4.0,0) {$\vdots$};

\node  at (0,0) {$\vdots$};
\node[draw, rectangle, minimum height=2.5cm, minimum width=2.7cm ](Plant)
at (-2,0) {$\mathrm{Network~A}$};

\node[draw, rectangle, minimum height=2.5cm, minimum width=2.7cm ](Plant)
at (1.85,0) {$\mathrm{Network~B}$};

\draw [->] (-4.5,1.05) -- (-3.3,1.05) {} ; 
\draw [-] (-4.1,0.65) -- (-3.4,0.65) {} ; 
 \draw [-] (-4.5,0.65) -- (-4.35,0.65) {} ; 
 
 \draw [-] (-4.5,-1.05) -- (-3.85,-1.05) {} ; 
\draw [->] (-4.5,-0.65) -- (-3.3,-0.65) {} ; 
 \draw [-] (-3.6,-1.05) -- (-3.4,-1.05) {} ;

\draw [<-] (3.2,1.05) -- (4.0,1.05) {} ; 
\draw [-] (3.2,0.65) -- (4.0,0.65) {} ;

 \draw [-] (3.2,-1.05) -- (4.0,-1.05) {} ; 
\draw [<-] (3.2,-0.65) -- (4.0,-0.65) {} ;

 \draw [-] (-3.6,-1.05) --(-3.6,-1.85)--(3.4,-1.85)--(3.4,-0.65){} ;
 
 \filldraw (3.4,-0.65) circle (1pt);
 
  \draw [-] (-3.85,-1.05) --(-3.85,-2.05)--(3.48,-2.05)--(3.48,-1.05){} ;
 
 \filldraw (3.48,-1.05) circle (1pt);

 \draw [-] (-4.35,0.65) --(-4.35,-2.65)--(3.85,-2.65)--(3.85,0.65){} ;
 \filldraw (3.85,0.65) circle (1pt);

 \draw [-] (-4.1,0.65) --(-4.1,-2.45)--(3.75,-2.45)--(3.75,1.05){} ;
  \filldraw (3.75,1.05) circle (1pt);
 \node  at (0,-2.15) {$\vdots$};
\draw [-] (-0.66,-1.05) -- (0.5,-1.05) {} ; 
\draw [<->] (-0.66,-0.65) -- (0.5,-0.65) {} ;

\draw [<->] (-0.66,1.05) -- (0.5,1.05) {} ; 
\draw [-] (-0.66,0.65) -- (0.5,0.65) {} ;  

 \node  at (-2.9,0.85) {{\footnotesize $\mathrm{port}~1$}};
 \node  at (-2.9,-0.85) {{\footnotesize $\mathrm{port}~r$}};
 \node  at (-1.35,0.85) {{\footnotesize $\mathrm{port}~r\!+\!1$}};
 \node  at (-1.35,-0.85) {{\footnotesize $\mathrm{port}~r\!+\!s$}};

 \node  at (2.55,0.85) {{\footnotesize $\mathrm{port}~s\!+\!1$}};
 \node  at (2.55,-0.85) {{\footnotesize $\mathrm{port}~s\!+\!r$}};
 \node  at (1.02,0.85) {{\footnotesize $\mathrm{port}~1$}};
 \node  at (1.02,-0.85) {{\footnotesize $\mathrm{port}~s$}};
\end{tikzpicture}
}
\end{center}
\caption{Hybrid-Cascade connection}
\label{fig:hybrid-cascade connection}
\end{figure}
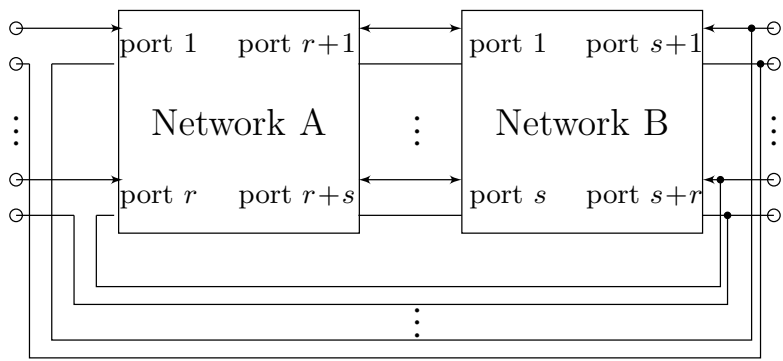

\textbf{(vii) Hybrid-Cascade connection:}
Assume that $A$ and $B$ are both $(r+s)$-port networks. 
The connection in Fig.~\ref{fig:hybrid-cascade connection}  combines the hybrid and cascade connections 
as
\begin{equation}
\begin{aligned}
&V_i^c=V_i^a+V_{s+i}^b, &I_i^c=I_i^a,~~~~~~~~~~~~&i=1,2,\ldots,r,
\\
&V_{r+i}^a=V_i^b,      &I_{r+i}^a=-I_i^b,~~~~~~~~ &i=1,2,\ldots,s,
\\
&V_{r+i}^c=V_{s+i}^b,  &I_{r+i}^c=I_{s+i}^b-I_{i}^a,~~ &i=1,2,\ldots,r.
\end{aligned}
\end{equation}
The explicit matrix representation of the resulting network is provided in Appendix A and is also denoted by
\begin{equation}
Z^c=Z^a \star Z^b.
\end{equation}

Note that these abbreviated symbols $\left\lbrace \vcentcolon, \ast,\circ,\star \right\rbrace$ are standard and frequently adopted in the pertinent literature ($\!\!$\cite{anderson1975matrix,mitra1975hybrid,duffin1978almost,anderson1986cascade,mitra1982shorted}).


Armed with the phase knowledge of $n$-port subnetworks, we extended the result in Proposition \ref{prop:1} and studied the phase properties of the obtained networks under the connections demonstrated above. 

\begin{theorem}\label{thm:phase-preserving}
If 
$
Z^a,~Z^b\in \mathcal{Z}[\alpha, \beta], 
$
it holds that
\begin{flalign}
&1)\qquad\qquad\qquad\quad\qquad\quad\qquad\quad Z^c= Z^a/_{22} \in \mathcal{Z}[\alpha, 
\beta],&\label{eq:phase-preserving 1}
\\
&2)\qquad\qquad\qquad\quad\qquad\quad\qquad\quad  Z^c= Z^a_{11} \in \mathcal{Z}[\alpha, 
\beta],&\label{eq:phase-preserving 2}
\\
&3)\qquad\qquad\qquad\quad\qquad\quad\qquad\quad Z^c= Z^a \# Z^b  \in \mathcal{Z}[\alpha,\beta],&\label{eq:phase-preserving 3}
\\
&\qquad\qquad\qquad\qquad\quad \qquad\qquad\qquad\qquad  \forall \# \in \left\lbrace +,  \vcentcolon,  \ast,\circ,\star \right\rbrace.&
\nonumber
\end{flalign}
\end{theorem}

The proof of this theorem is given in Appendix B.
As evident, phase-preserving properties \eqref{eq:phase-preserving 1}, \eqref{eq:phase-preserving 2}, and \eqref{eq:phase-preserving 3} can be naturally extended to large-scale complex networks with multiple phase-bounded subnetworks under arbitrary connections aforementioned. A \textit{scalable} phase-preserving criterion can be proposed provided the overall phases of all the subnetworks are contained in the same convex cone at each frequency.
This criterion can simplify  the analysis of 
phase-related properties of large-scale networks by examining their relatively low-order and comprehensible
subnetworks. 
To facilitate this understanding, we 
consider three basic one-port networks: a resistor with $Z^R=R$, a capacitor with $Z^C=1/(j\omega C)=e^{-j \frac{\pi}{2}}/(\omega C)$, and an inductor with
$Z^L=j\omega L=\omega L e^{j\frac{\pi}{2}}$.
Therefore, based on the results above, regardless of an arbitrarily intricate one-port network comprising only resistors, capacitors, and inductors, the phases are confined to the closed right-half complex plane.
Apparently, such a phase relation can be extended to the $n$-port networks interconnected by $n$-port resistive, capacitive, and inductive networks, 
implying  the well-known passivity-preserving result \eqref{eq:passive 2} for networks.
Thus, we firmly state that our results quantify the passivity of networks in the form of phases. Moreover,
the passivity-preserving result was generalized to a wider range of passive or active networks in a frequency-wise cone-bounded manner.

\begin{remark}\label{remark:1}
As discussed in the shorted connection, the existence of $Z^c$ can also be guaranteed under other connections \textbf{(iii)–(vi)}, if certain conditions, such as those described in \eqref{eq:well-posedness}, are satisfied. Further details regarding the presence of $Z^c$ for general connections are presented in Section~\ref{sec:Extension}. Note that, for example, it has been shown in \cite{anderson1975shorted} that if $Z^a$ and $Z^b$ represent resistive networks, then $Z^c$ always exists. More generally, consider passive networks $Z^a$ and $Z^b$ that satisfy an improved positive real condition, given by $\mathrm{Re}\left(\left\langle Zx,x\right\rangle\right)=\mathrm{Re}\left(x^\ast Zx\right) \geq 0$ for all $x\in \mathbb{C}^n$, and $\mathrm{Re}\left(\left\langle Zx,x\right\rangle\right)=0$ only if $Zx=0$. Then $Z^c$ always exists (see \cite{mitra1982shorted,mitra1975hybrid,anderson1986cascade,mitra1986parallel,mitra1983fundamental}).
\hfill $\blacksquare$
\end{remark}

\begin{remark}
All the connections of $n$-port networks addressed previously are only valid under the assumption that ideal isolation transformers are placed to ensure proper port behaviors.
Even the series connection formula in \eqref{eq:Zc Series connection} fails without these transformers.
\hfill $\blacksquare$
\end{remark}

\begin{remark}
There exists another celebrated connection, the feedback connection \cite{zhou1996robust}, which is considered more common in control systems than in electrical networks. The relationship between the phases of the open-loop system and that of the associated closed-loop system under feedback connection has been previously revealed
in \cite{chen2021phase,mao2022phases}.
\hfill $\blacksquare$
\end{remark}

\section{Subtractions of Networks}\label{sec: Subtraction}
Subtractions are inverse operations to network connections in a one-to-one match. For example, 
for a series connection involving two networks $A$ and $B$, the obtained network $C$ can be characterized with $Z^c=Z^a+Z^b$. The inverse operation is used to study network $A$ with the known networks $B$ and $C$ only.
We use $X$ to function as an unknown network. Consequently, it can be observed that:

\textbf{(i) Series subtraction}
\begin{equation}\label{eq:Zx Series subtraction}
Z^x=Z^c-Z^b.
\end{equation}
Therefore, parallel, hybrid, cascade, and hybrid-cascade subtractions can be defined under similar rules with respect to their corresponding connections and are abbreviated as follows (the explicit formulae of $Z^x$ are provided in Appendix A):

\textbf{(ii) Parallel subtraction}
\begin{equation}\label{eq:Zx Parallel subtraction}
Z^x=Z^c \vcentcolon\inv Z^b=-Z^b(Z^c-Z^b)^{-1}Z^c.
\end{equation}

\textbf{(iii) Hybrid subtraction}
\begin{equation}\label{eq:Hybrid Zx1}
Z^x=Z^c \ast\inv Z^b.
\end{equation}

\textbf{(iv) Cascade subtraction}
\begin{equation}\label{eq:Cascade Zx1}
Z^x=Z^c \circ\inv Z^b.
\end{equation}

\textbf{(v) Hybrid-Cascade subtraction}
\begin{equation}\label{eq:Hybrid-Cascade Zx1}
Z^x=Z^c \star\inv Z^b.
\end{equation}

The 
network subtractions pertain to the matrix equation-solving problems. The existence of such solutions $Z^x$ should be confirmed under certain conditions, which have been
provided in \cite{anderson1986cascade}, and are omitted herein because of their tediousness. Moreover, it can be verified that such conditions can always be satisfied by frequency-wise sectorial networks. 
Those corresponding subtraction operations could be interpreted in terms of the synthesis of $n$-port networks: Given a specified $n$-port network, determine the set of $n$-port networks which, when connected with the specified network, yield another $n$-port network satisfying certain prescribed requirements. Clearly, the phase can be used as the limitation or requirement of networks to address subtraction operations more fully.
Following a trend similar to Theorem~\ref{thm:phase-preserving}, 
one natural question is, in virtue of the phase bounds of $Z^b$ and $Z^c$, what can be the feasible phase range of $Z^x$? Again, as in the role of Proposition~\ref{prop:1}, first, certain insights were obtained from the SISO cases.
Consider the frequency responses of three one-port networks, denoted by $z^x=\sigma^x e^{j\theta^x}$, $z^b=\sigma^b e^{j\theta^b}$, and $z^c=\sigma^c e^{j\theta^c}$.

 \begin{proposition}\label{prop:2}
 Let $z^x=z^c-z^b.$ 
If $\theta^c-\theta^b\in [0,2\pi)$, 
  \begin{equation}\label{eq:phase relationship SISO subtracton1}
\theta^x\! \in \!\left[\min\left\lbrace  \theta^c,\theta^b\!+\!\pi  \right\rbrace\!,  \max\left\lbrace  \theta^c,\theta^b\! +\!\pi \right\rbrace   \right];
 \end{equation}
If  $\theta^c-\theta^b\in [-2\pi, 0)$, 
  \begin{equation}\label{eq:phase relationship SISO subtracton2}
  \theta^x \!\in \!\left[\min\left\lbrace  \theta^c,\theta^b\!-\!\pi  \right\rbrace\!,  \max\left\lbrace  \theta^c,\theta^b\!\!-\pi  \right\rbrace   \right].
 \end{equation}
 \end{proposition}

Now, let us consider two $n$-port networks, $B$ and $C$ with $Z^{b} \in \mathcal{Z}[\alpha^{b}, \beta^{b}]$ and $Z^{c} \in \mathcal{Z}[\alpha^{c}, \beta^{c}]$.
The preceding results \eqref{eq:phase relationship SISO subtracton1}-\eqref{eq:phase relationship SISO subtracton2} can be generalized to $n$-port cases as follows, where arbitrary subtractions 
were considered in the unified criterion.

\begin{theorem}\label{thm:phase-preserving2}
If $[\alpha^{b}, \beta^{b}]\cap [\alpha^{c}, \beta^{c}]=\emptyset$ mod $2\pi$, 
 it holds that
\begin{equation} \label{eq:thm2 condition}
\begin{aligned}
&Z^{x}=
Z^c \# Z^b 
 \in\! \mathcal{Z}[\min\lbrace  \alpha^{c},   \alpha^{b}\pm\pi    \rbrace ,  \max\lbrace   \beta^{c},   \beta^{b}\pm\pi    \rbrace ], 
\\
&~~~~~~~~~~~~ \qquad\qquad\qquad  \forall \#  \in \left\lbrace -,  \vcentcolon\inv,  \ast\inv,\circ\inv,\star\inv \right\rbrace,
\end{aligned}
\end{equation}
where $\pm$ is  determined by enabling $\max\lbrace   \beta^{c},   \beta^{b} \pm\pi  \rbrace-\min\lbrace   \alpha^{c},   \alpha^{b} \pm\pi    \rbrace<\pi$.

\end{theorem}

\begin{figure}[h]
\begin{center}
\includegraphics[width=6.2cm, height=6.4cm]{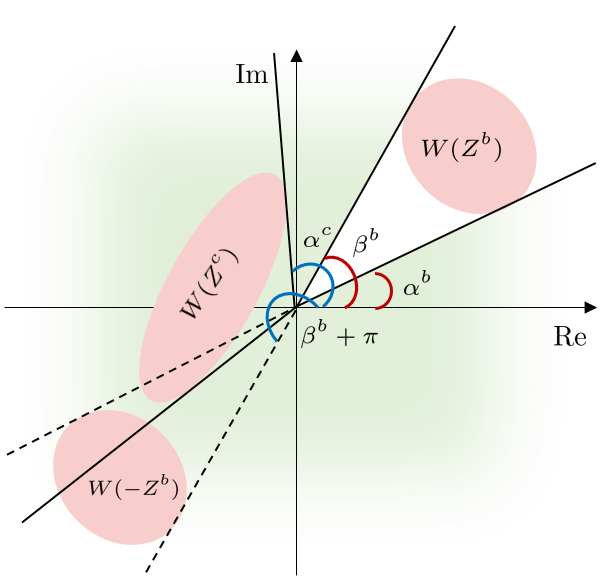}
\end{center}
\caption{Phase ranges of $Z^{b}$, $Z^{c}$, and  $Z^{x}$ at a fixed $\omega$.}
\label{fig:Phase ranges subtraction}
\end{figure}

The proof is given in Appendix C.
Similar to the connections, the phase range in \eqref{eq:thm2 condition} is independent of the choice of subtractions.
Figure \ref{fig:Phase ranges subtraction} provides an explanation for Theorem \ref{thm:phase-preserving2}.
Thus, using the numerical range of $Z^b$ at a fixed frequency $\omega$,  the corresponding phase range $[\alpha^{b}, \beta^{b}]$
can be determined. The aqua region characterizes the maximal allowable phase range for $Z^c$ such that 
$[\alpha^{b}, \beta^{b}]\cap [\alpha^{c}, \beta^{c}]=\emptyset$. As depicted in Fig.~\ref{fig:Phase ranges subtraction}, the explicit phase ranges of $Z^c$ and $-Z^b$ codetermine that of $Z^x$; that is, $Z^x\in \mathcal{Z}[\alpha^{c}, \beta^{b}+\pi]$.

\section{Extensions: Confluences and Operations}\label{sec:Extension}
Although fundamental limitations exist in series and parallel connections,
most classical studies on connections of $n$-port networks have been restricted to series and parallel structures ($\!\!$\cite{shannon1949synthesis,anderson1969series,steigerwald1988comparison}).
However, other more complex connections offer enormous advantages over several sophisticated network problems. 
 For instance, it was found that the hybrid connection can considerably simplify network realization and synthesis problems \cite{willems1976realization,duffin1966network}.
This status quo motivated us to study more general network connections by incorporating
the cascade and hybrid cascade connections. However, the question is whether the connections considered above are sufficiently general. Further, do all the potential connections possess intrinsic phase-preserving properties? To address these questions, in this section, we need further elaboration. 

Given an arbitrary connection, the only required condition that must be satisfied is Kirchhoff's laws, which can be characterized by the relations of currents (voltages). Therefore, all network connections shall imply specified relations defined in the space of current (voltage) vectors.
To this end,
we begin by introducing the notion of confluence, a subspace of a 3$n$-dimensional space admitting certain axioms.
\begin{defn} ($\!\!$\cite{anderson1975matrix})
A subspace $\textbf{G}$ of $\textbf{V}=\mathbb{R}^n \times \mathbb{R}^n \times \mathbb{R}^n$ is referred to as a confluence if:
\begin{enumerate}
\item[(i)] if $(0,0,c)$ is in $\textbf{G}$, then $c=0$.

\item[(ii)] for each vector $c\in \mathbb{R}^n$, there is a pair $(a,b)$ such that $(a,b,c)$ is in $\textbf{G}$.
\end{enumerate}

\end{defn}
Vector $x$ in $\textbf{V}$ is written $x=(a,b,c)$, where $a,b,c\in \mathbb{R}^n$. We define an indefinite inner product as 
 $\left[ \cdot \; , \; \cdot \right] : \textbf{V} \times \textbf{V} \rightarrow \mathbb{R} $
by the formula:
\begin{equation}
\label{eq:indefinite inner product}
\left[ x,y\right] =\left[ (a_1,b_1,c_1),(a_2,b_2,c_2)\right] =\langle a_1 , a_2 \rangle  +\langle b_1 , b_2 \rangle  -\langle c_1 , c_2 \rangle, 
\end{equation}
equipped with $\langle a_1 , a_2 \rangle = a_1^{\ast}a_2$. 
Based on such an indefinite inner product, the orthogonal complement of confluence $\textbf{G}$, denoted by $\textbf{G}^{\perp}\subset \textbf{V}$, can be defined 
as
\begin{equation}
\textbf{G}^{\perp}:=\left\lbrace y: \left[ x,y \right]=0,  \forall x \in \textbf{G} \right \rbrace . 
\end{equation}
Clearly, $\textbf{G}^{\perp \perp}=\textbf{G}$ holds. 
We also have $\mathrm{dim}~\textbf{G}+\mathrm{dim}~\textbf{G}^{\perp}=\mathrm{dim} \textbf{V}=3n.$  
It has been proven in \cite{anderson1975matrix} that $\textbf{G}^{\perp}$ is also a confluence. 
Then from the definition of confluence and dimension equality above, we have 
\begin{equation}
n\leq \mathrm{dim}~\textbf{G} \leq 2n.
\end{equation}
Physically, the reason for introducing $\textbf{G}$ and $\textbf{G}^{\perp}$ is that, if $\textbf{G}$ is the set of all possible current vectors for
a network connection, then the set $\textbf{G}^{\perp}$ contains all the possible voltage vectors. 

By choosing the matrices whose columns span $\textbf{G}$ and $\textbf{G}^{\perp}$ properly,
the confluences can be represented in terms of the matrices.

 \begin{proposition}($\!\!\!~$\cite{anderson1975matrix}) \label{prop:confluence matrix representation}
Given a confluence $\textbf{G}$, for all vectors $(a,b,c)\in \textbf{G}$, there exist the matrices 
$S, T, U$, and $W\in \mathbb{R}^{n\times n}$ such that
\begin{equation}\label{eq:G confluence}
\begin{bmatrix}
S&T\\
U&W
\end{bmatrix}
\begin{bmatrix}
a\\b
\end{bmatrix}=
\begin{bmatrix}
c
\\0
\end{bmatrix}.
\end{equation}
For all vectors $(\alpha,\beta,\gamma)\in \textbf{G}^{\perp}$, there exist the matrices 
$\Phi, \Psi, \Xi$, and $\Omega\in \mathbb{R}^{n\times n}$ such that
\begin{equation}\label{eq:confluence four matrices}
\begin{bmatrix}
\Phi &\Psi
\\
\Xi&\Omega
\end{bmatrix}
\begin{bmatrix}
\alpha \\ \beta
\end{bmatrix}=
\begin{bmatrix}
\gamma
\\0
\end{bmatrix}.
\end{equation}
Those matrices satisfy
\begin{equation}
\begin{bmatrix}
S&T\\
U&W
\end{bmatrix}
\begin{bmatrix}
\Phi^{\ast} & \Xi^{\ast} 
\\
\Psi^{\ast}&\Omega^{\ast}
\end{bmatrix}
=
\begin{bmatrix}
I&0\\
0&0\\
\end{bmatrix}.
\end{equation}

 \end{proposition}
 
Note that given a confluence $\textbf{G}$, its matrix representation in \eqref{eq:G confluence}  may not be unique as long as the null space of the matrix
\begin{equation}
\begin{bmatrix}
S&T&-I\\
U&W&0
\end{bmatrix}
\nonumber
\end{equation}
coincides with $\textbf{G}$.
 \begin{proposition}
The set of all  matrix representations of $\textbf{G}$ and $\textbf{G}^{\perp}$ can be parametrized by
\begin{equation}\label{eq:confluence parameterization}
\begin{bmatrix}
I&P\\
0&Q
\end{bmatrix}
\begin{bmatrix}
S&T
\\
U&W
\end{bmatrix}
=
\begin{bmatrix}
S+PU&T+PW
\\
QU&QW
\end{bmatrix},
\end{equation}
and
\begin{equation}\label{eq:confluence perp parameterization}
\begin{bmatrix}
I&\Gamma\\
0&\Lambda
\end{bmatrix}
\begin{bmatrix}
\Phi &\Psi
\\
\Xi&\Omega
\end{bmatrix}
=
\begin{bmatrix}
\Phi+\Gamma \Xi&\Psi+\Gamma\Omega
\\
\Lambda\Xi&\Lambda\Omega
\end{bmatrix},
\end{equation}
for all $P,Q,\Gamma,\Lambda\in \mathbb{R}^{n\times n}$ and $\mathrm{Rank}~(Q)=\mathrm{Rank}~(\Lambda)=n.$ 
 \end{proposition}

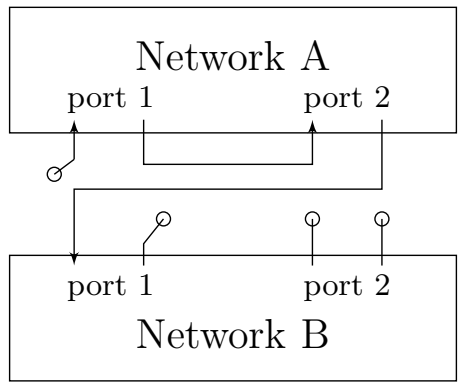
\begin{figure}
\begin{center}
 \resizebox{0.350\columnwidth}{!}{
\begin{tikzpicture}[auto, node distance=3cm,>=latex']


\draw (-1.8,1.45) circle (2pt);
\draw (-0.7,1.0) circle (2pt);

\draw (0.8,1.0) circle (2pt);
\draw (1.5,1.0) circle (2pt);


\node[draw, rectangle, minimum height=3em, minimum width=4.5cm ](Plant)
at (0,0) {};
\node  at (0,-0.15) {$\mathrm{Network~B}$};
 
\node[draw, rectangle, minimum height=3em, minimum width=4.5cm ](Plant)
at (0,2.5) {};
\node  at (0,2.65) {$\mathrm{Network~A}$};

\draw [-] (-1.8,1.45)--(-1.6,1.6){}; 
\draw [->](-1.6,1.6)--(-1.6,2.0){}; 

\draw [-] (-0.9,0.53)--(-0.9,0.75) -- (-0.7,1.0) {}; 
 

\draw [->] (-0.9,2)--(-0.9,1.55)--(0.8,1.55)--(0.8,2) {}; 

\draw [->] (1.5,2)--(1.5,1.3)--(-1.6,1.3)--(-1.6,0.53) {}; 

\draw [-] (0.8,0.53)--(0.8,1) {}; 
 
\draw [-] (1.5,0.53)--(1.5,1) {};

 
 \node  at (-1.23,0.3) {{\footnotesize $\mathrm{port}~1$}};
 \node  at (1.15,0.3) {{\footnotesize $\mathrm{port}~2$}};
 
 \node  at (-1.23,2.2) {{\footnotesize $\mathrm{port}~1$}};
 \node  at (1.15,2.2) {{\footnotesize $\mathrm{port}~2$}};
 
\end{tikzpicture}
}
\end{center}
\caption{A new type of connection}
\label{fig:new connection}
\end{figure}

The proof is straightforward and hence is omitted for brevity.
In the preceding sections, we studied the network connections
arising from the graphical connections of the set of terminals (Figs. \ref{fig:shorted connection}-\ref{fig:hybrid-cascade connection}).
Such a matrix treatment is considerably more general, and it has been shown in \cite{anderson1975matrix} that all physically realizable network
connections can be explained by certain confluences and not vice versa.
Specifically, consider a network connection equipped with confluence $\textbf{G}$, and 
 let the currents of networks $A$ and $B$ and that of the integrated network $C$ 
 satisfy $(I^a, I^b, I^c) \in \textbf{G}$. Then, the corresponding voltages must satisfy
 $(V^a, V^b, V^c) \in \textbf{G}^{\perp}$.
 For example, for a series connection, we have the commonly seen representations
 \begin{equation}
\begin{bmatrix}
 S &T\\
 U&W
 \end{bmatrix}= \begin{bmatrix}
I&0\\
I&-I
 \end{bmatrix},
~~~\begin{bmatrix}
\Phi &\Psi
\\
\Xi&\Omega
\end{bmatrix}
=
\begin{bmatrix}
I&I\\
0&0
 \end{bmatrix}.
\end{equation}
Matrices defining the parallel connections can be expressed as
\begin{equation}
 \begin{bmatrix}
 S &T\\
 U&W
 \end{bmatrix}= \begin{bmatrix}
I&I\\
0&0
 \end{bmatrix},
~~~
\begin{bmatrix}
\Phi &\Psi
\\
\Xi&\Omega
\end{bmatrix}
=
\begin{bmatrix}
I&0\\
I&-I
 \end{bmatrix}.
\end{equation}
The hybrid, cascade, and hybrid-cascade network connections  studied in Section \ref{sec:connections} can be also 
characterized by such associated matrices, respectively. 
Next, we utilize a specified confluence to illustrate an exactly new type of connection between two 2-port networks.
Such a confluence holds the matrix representations:
 \begin{equation}\label{exam:new connection}
\begin{bmatrix}
 S &T\\
 U&W
 \end{bmatrix}\!=\! \begin{bmatrix}
1&0&0&0\\
0&0&0&1
\\
1&-1&0&0\\
1&0&-1&0
 \end{bmatrix},
~~\begin{bmatrix}
\Phi &\Psi
\\
\Xi&\Omega
\end{bmatrix}
\!=\!
\begin{bmatrix}
1&1&1&0\\
0&0&0&1\\
0&0&0&0
\\
0&0&0&0
 \end{bmatrix}.
\end{equation}
The corresponding graphical illustration is given in Fig.~\ref{fig:new connection}.

We enabled the unification and extension of the connections with the aid of confluence: 
For an arbitrary connection associated with a pair of dual confluences $\textbf{G}$ and $\textbf{G}^{\perp}$, along with two $n$-port networks $A$ and $B$ having $Z^a$ and $Z^b$ respectively, we examine  $Z^c$ of the integrated $n$-port network $C$. To clarify the following theorem more clearly, we first define the matrix 
\begin{flalign}
M&=\begin{bmatrix}\label{eq:Zc general}
\Phi&\Psi\\
\Xi&\Omega
\end{bmatrix} 
\begin{bmatrix}
Z^a&0\\
0&Z^b
\end{bmatrix} 
\begin{bmatrix}
\Phi^{\ast}&\Xi^{\ast}\\
\Psi^{\ast}&\Omega^{\ast}
\end{bmatrix}, 
\end{flalign}
and $\Phi, \Psi, \Xi$, and $\Omega$ are presented in \eqref{eq:confluence four matrices}. We show that $Z^c$ can be compactly represented by the Schur complement of  $M$, as elaborated below.

\begin{theorem}\label{thm: confluence representation}
The Z-matrix $Z^c$ exists if and only if $M/_{22}$ is well-defined. In such a case,
\begin{flalign}
Z^c&=M/_{22}.\label{eq:Zc}
\end{flalign}
\end{theorem}

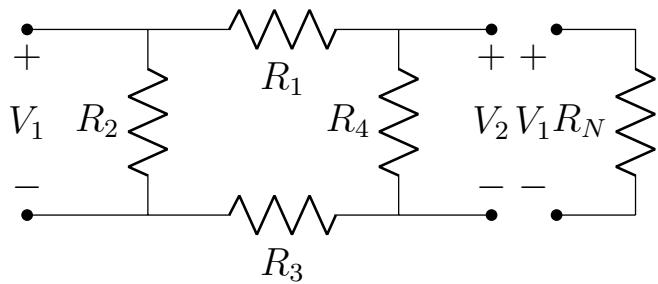
\begin{figure}
\begin{center}
 \resizebox{0.5\columnwidth}{!}{
\begin{circuitikz}[american voltages]
\draw

  (0,0) to [short, *-] (2.0,0)
  (3.5,0) to [short, -*] (5,0)
  
  (5.7,0) to [short, *-] (6.55,0)
   
  (0,-2) to [short, *-] (2,-2)
  (3.5,-2) to [short, -*] (5,-2)
  
  (5.7,-2) to [short, *-] (6.55,-2) 
  
  (2,0) to [R, l_=$R_1$] (3.5,0)
  (2,-2) to [R, l_=$R_3$] (3.5,-2)
  (1.3,0) to [R, l_=$R_2$] (1.3,-2)
  (4,0) to [R, l_=$R_4$] (4,-2)
  (6.55,0) to [R, l_=$R_N$] (6.55,-2)
  
(0,0) to [open, v^>=$V_1$] (0,-2)

(5.45,0) to [open, v^>=$V_1$] (5.45,-2) 
 
(5,0) to [open, v^>=$V_2$] (5,-2)
 ;
\end{circuitikz}
}
\end{center}
\caption{Non-existence of Z-matrix under cascade connection.}
\label{fig:resistive network}
\end{figure}

One may refer to Appendix D for the proof. 
As discussed in Section~\ref{sec:connections}, the well-definedness of $M/_{22}$ can be established by examining the condition \eqref{eq:well-posedness} accordingly, as demonstrated by the relationships \eqref{eq:Zc existence condition1}--\eqref{eq:Zc existence condition2} in Appendix~D. To illustrate this, consider two networks $A$ and $B$ with Z-matrices $Z^a$ and $Z^b$, respectively. When these networks are connected in cascade to form a new network $C$, it is possible that $C$ does not have a Z-matrix representation, as shown in Fig.~\ref{fig:resistive network}.
The left two-port network comprises four resistors $R_{i}, i=1,\ldots,4$ with the constant Z-matrix
$$Z^a=\begin{small}
\frac{\begin{bmatrix}
(R_1+R_3+R_4)R_2&R_2R_4
\\
R_2R_4&(R_1+R_2+R_3)R_4\end{bmatrix}}{R_1+R_2+R_3+R_4}
\end{small}
,
$$
while the right one-port network contains a static negative resistance that can be chosen as  $Z^b=R_N=-(R_1+R_2+R_3)R_4/(R_1+R_2+R_3+R_4)$.
It can readily be verified that the existence condition cannot be satisfied due to $Z^{a}_{22}+Z^b=0$, so $Z^c$ does not exist.

After confirming the existence of $Z^c$, the formula  \eqref{eq:Zc} generalizes all explicit expressions of $Z^c$ under distinct connections (e.g., \eqref{eq:Hybrid Zc2}–\eqref{eq:Hybrid-Cascade Zc2}).
Under such a unification of connections,
the phase-preserving property of $Z^c$ with respect to $Z^a,~Z^b\in \mathcal{Z}[\alpha, \beta]$ can be further studied.

\begin{theorem}\label{thm:phase-preserving confluence}
For an arbitrary connection, if $Z^a, Z^b\in \mathcal{Z}[\alpha,  \beta]$, then $Z^c$ in \eqref{eq:Zc} satisfies
\begin{equation}
Z^c\in \mathcal{Z}[\alpha, \beta].
\end{equation}

\end{theorem}

The proof is given in Appendix E.
Theorem \ref{thm:phase-preserving confluence} indicates the most general result regarding the phase range of $n$-port networks under connections: 
\begin{align*}
    \textit{All possible connections preserve the phase}. 
\end{align*}

Previous research has focused primarily on the preservation of passivity in $n$-port networks under connections of interest. 
For example, it is shown in \cite{anderson2013network} that if $A$ and $B$ are passive networks, the resultant series-, parallel-, and cascade-connected networks $C$ are still passive.
The relationship between the phase and passivity has been discussed in Section \ref{subsec:passive network}. 
Passivity is a qualitative description of networks that can be quantified by the phase acting as a measure of the surplus of passivity. Furthermore, the phase can be used to characterize certain non-passive networks meeting the so-called semi-sectorial property, which extends the scope of passivity from a phasic perspective. For instance, if the maximal phase of a network over all frequencies is slightly greater than $\pi/2$, the network is not passive. However, in terms of phase, we can say that the network is ``barely'' passive, and its behavior observed at ports shall have a certain similarity with that of passive networks. Of course, such a bare passivity can be preserved independent of all possible connections. 
In summary, Theorem~\ref{thm:phase-preserving confluence} fundamentally generalizes the classical passivity-preserving results from a viewpoint of phase.

\section{Illustrative Example}\label{sec: example}

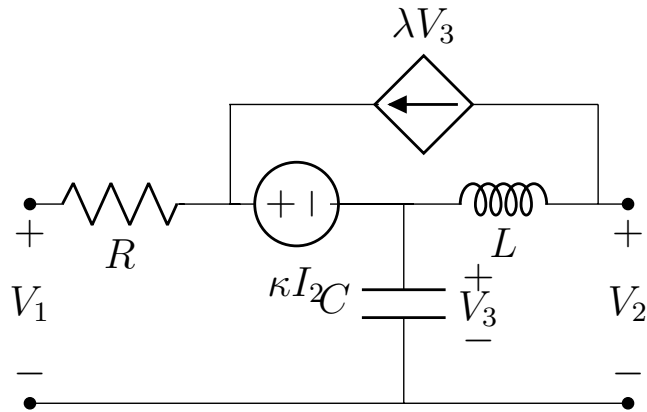
\begin{figure}
\begin{center}
 \resizebox{0.5\columnwidth}{!}{
\begin{circuitikz}[american voltages]
\draw
  (0,0) to [short, *-] (0.25,0)
  (2.0,0) to [short, -] (2.11,0)
  (3.01,0) to [short, -] (4,0)
  (5.5,0) to [short, -*] (6,0)
  (0,-2) to [short, *-*] (6,-2)
  (2.01,0) to [short, -] (2.01,1)
  (5.7,0) to [short, -] (5.7,1)
   (4.45,1) to [short, -] (5.7,1)
   (2,1) to [short, -] (3.5,1)
  (0,0) to [open, v^>=$V_1$] (0,-2) 
  (4.5,-0.6) to [open, v^>=$V_3$] (4.5,-1.5) 
  (6,0) to [open, v^>=$V_2$] (6,-2)
  (0.25,0) to [R, l_=$R$] (1.55,0)
  (3.75,0) to [C, l_=$C$] (3.75,-2)
  (4,0) to [L, l_=$L$] (5.5,0)
(1.5,0) to  [american controlled  voltage  source,  V_=$\kappa I_2$](3.85,0)
(4.45,1) to [american controlled current source, l_=$\lambda V_3$] (3.45,1)
 ;
\end{circuitikz}
}
\end{center}
\caption{Another two-port network.}
\label{fig:B 2-port network}
\end{figure}

\begin{figure}[h]
\begin{center}

\includegraphics[width=10cm, height=3.25cm]{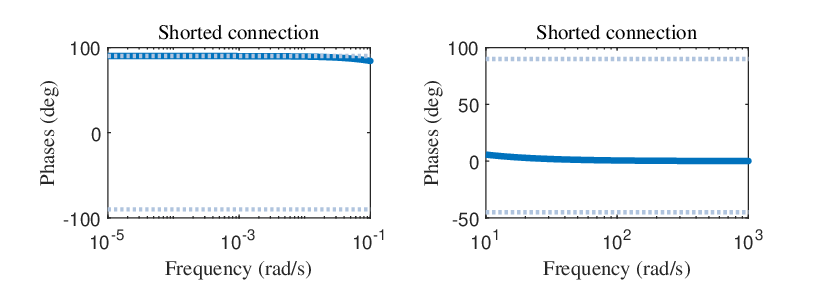}
\end{center}
\begin{center}
\vspace{-0.5cm}
\includegraphics[width=10cm, height=3.25cm]{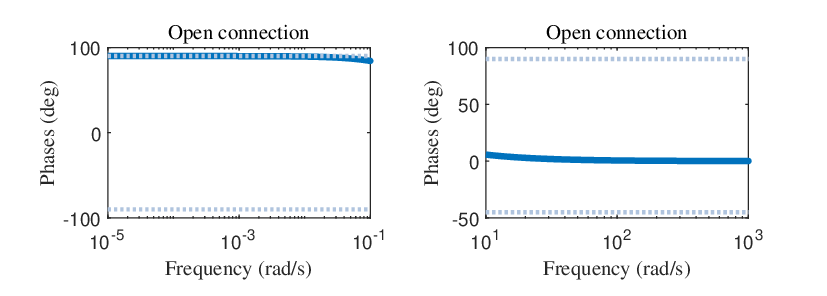}
\end{center}
\begin{center}
\vspace{-0.5cm}
\includegraphics[width=10cm, height=3.25cm]{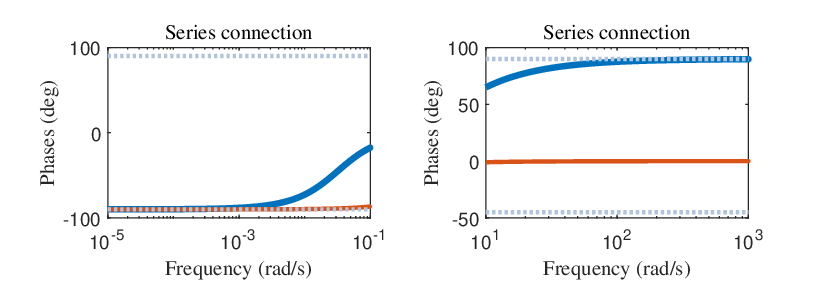}
\end{center}
\begin{center}
\vspace{-0.5cm}
\includegraphics[width=10cm, height=3.25cm]{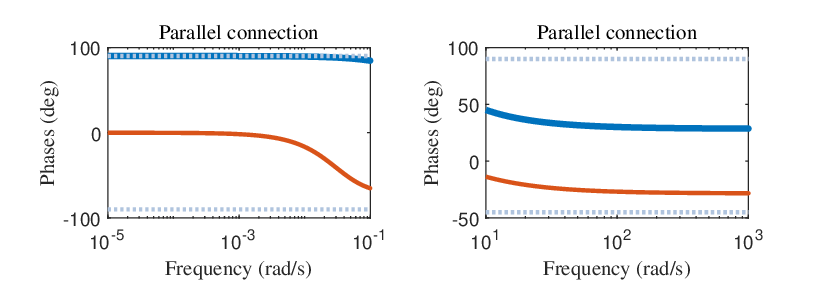}
\end{center}
\begin{center}
\vspace{-0.5cm}
\includegraphics[width=10cm, height=3.25cm]{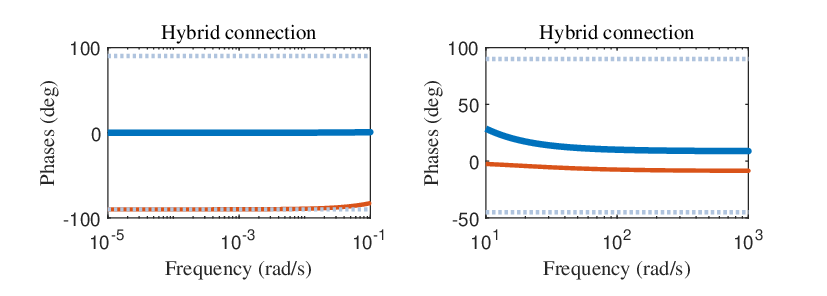}
\end{center}
\begin{center}
\vspace{-0.5cm}
\includegraphics[width=10cm, height=3.25cm]{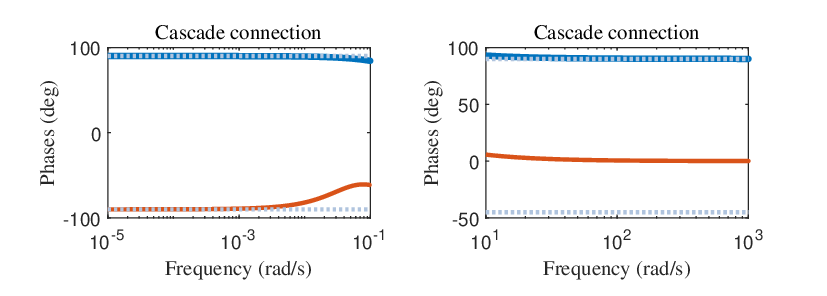}
\end{center}
%
\begin{center}
\vspace{-0.5cm}
\includegraphics[width=10cm, height=3.25cm]{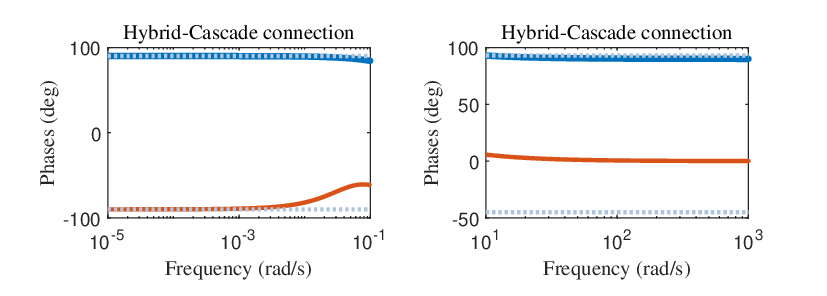}
\end{center}
\caption{Phase ranges of $Z^c$ under enumerative connections: $Z^a/_{22}$, $ Z^a_{11}$, and $ \left\lbrace +, \vcentcolon, \ast,\circ,\star \right\rbrace$ (left $\omega\in [0,10^{-1}]$ and right $\omega\in [10^1,10^{3}]$).}
\label{fig:Phase ranges example}
\end{figure}

To illustrate the preceding phase-preserving results using numerical examples, we considered two $2$-port networks and examined their frequency-wise phase ranges
before and after the connections. A $2$-port network is shown in Fig.~\ref{fig:A 2-port network} whose Z-matrix is expressed in \eqref{eq:2port} with $R=1$, $L=1$, $C=2$, and $\gamma=1$. We denote this as $Z^a$ satisfying
\begin{equation}
Z^a(j\omega)\in \left\lbrace 
\begin{aligned}
&\mathcal{Z}[-\pi/2,  \pi/2],\qquad \forall \omega\in [0,10^{-1}],
\\
&\mathcal{Z}[-\pi/4,   \pi/4],\qquad \forall \omega\in [10,\infty].
\end{aligned}
\right. 
\nonumber
\end{equation}
Another $2$-port network is shown in Fig.~\ref{fig:B 2-port network} with its Z-matrix representation:
\begin{equation}
Z^b(s)=
\dfrac{1}{Cs}
\begin{bmatrix}
1+RCs&1+\kappa Cs
\\
1-\lambda Ls&1-\lambda Ls+LCs^2
\end{bmatrix}.
\end{equation}
Let $R=10$, $L=0.2$, $C=1$, $\kappa=0.1$, and $\lambda=0.5$. The phases of $Z^b$ are bounded by:
\begin{equation}
Z^b(j\omega)\in \left\lbrace 
\begin{aligned}
&\mathcal{Z}[-\pi/2,  0],\qquad \forall \omega\in [0,10^{-1}],
\\
&\mathcal{Z}[~0,  \!~~ \pi/2],\qquad \forall \omega\in [10,~~\!\infty].
\end{aligned}
\right. 
\nonumber
\end{equation}
Therefore, according to the phase-preserving criterion in Theorem \ref{thm:phase-preserving},
the integrated network shall fulfil
\begin{equation}
Z^c(j\omega)\in \left\lbrace 
\begin{aligned}
&\mathcal{Z}[-\pi/2,  \pi/2],\qquad \forall \omega\in [0,10^{-1}],
\\
&\mathcal{Z}[-\pi/4,  \pi/2],\qquad \forall \omega\in [10,~~\!\infty].
\end{aligned}
\right. 
\nonumber
\end{equation}
 This should be independent of the choice of connections, just as illustrated in Fig.~\ref{fig:Phase ranges example} (the end of the article).

\section{Conclusion}\label{sec: conclusion}
Inspired by the relatively new phase definition of complex
matrices based on the Z-matrix of an $n$-port network,
this study defined the frequency-wise phases of the network
accordingly. For a class of phase-bounded subnetworks, 
it was revealed that the integrated network involving the considered connections possessed phase-preserving properties. Thus, the
phase range could be determined effectively using
a scalable criterion. In addition, the results obtained could be
used to incorporate and quantify the proverbial passivity-preserving
properties of the connections of passive networks. 
We also studied network subtraction as an inverse operation of 
the connection. Therefore, the phase range of the obtained network after subtraction could be 
determined explicitly using a unified criterion. 
These classical network connections and associated phase analyses were extended
to a class of general matrix operations in a unified framework defined by the notion of confluence; that is a 3$n$-dimensional subspace
adopting an indefinite inner product. Under such a framework, we finally arrive at a crucial conclusion: All connections preserve the phase.

\section*{Appendix A}\label{appendix connection}
We collected the explicit expressions of $Z^c$ for hybrid, cascade, and hybrid-cascade connections.

\noindent
\textbf{Connections}

\textit{Hybrid connection:}
The partition \eqref{eq:Partition Z} results in natural partitions of $Z^b$ and $Z^c$.   The dimensions of the blocks are consistent with the indicated partition:
\begin{equation}
\begin{aligned}
Z^b&=
\begin{bmatrix}
Z^b_{11}&Z^b_{12}
\\
Z^b_{21}&Z^b_{22}
\end{bmatrix}\in \mathcal{R}^{\left( r+(n-r)\right) \times \left( r+(n-r)\right) },
\\
Z^c&=
\begin{bmatrix}
Z^c_{11}&Z^c_{12}
\\
Z^c_{21}&Z^c_{22}
\end{bmatrix}\in \mathcal{R}^{\left( r+(n-r)\right) \times \left( r+(n-r)\right) },
\end{aligned}
\nonumber
\end{equation}
where 
\begin{equation}\label{eq:Hybrid Zc2}
\begin{aligned}
Z^c_{11}&=Z^a_{11}+Z^b_{11}-(Z^a_{12}-Z^b_{12})(Z^a_{22}+Z^b_{22})^{\dagger}(Z^a_{21}-Z^b_{21}),\\
Z^c_{12}&=Z^a_{12}-(Z^a_{12}-Z^b_{12})(Z^a_{22}+Z^b_{22})^{\dagger}Z^a_{22},\\
Z^c_{21}&=Z^a_{21}-Z^a_{22}(Z^a_{22}+Z^b_{22})^{\dagger}(Z^a_{21}-Z^b_{21}),\\
Z^c_{22}&=Z^a_{22}(Z^a_{22}+Z^b_{22})^{\dagger}Z^b_{22}.\\
\end{aligned}
\end{equation}

\textit{Cascade connection:}
The partitions $Z^a$, $Z^b$, and $Z^c$ are compatible. 
It follows that $Z^a\in \mathcal{R}^{\left( r+s\right) \times \left( r+s\right) }$,  $Z^b\in \mathcal{R}^{\left( s+t\right) \times \left( s+t\right)}$,  $Z^c\in \mathcal{R}^{\left( r+t\right) \times \left( r+t\right) }$,
where
\begin{equation}\label{eq:Cascade Zc2}
\begin{aligned}
Z^c_{11}&=Z^a_{11}-Z^a_{12}(Z^a_{22}+Z^b_{11})^{\dagger}Z^a_{21},\\
Z^c_{12}&=Z^a_{12}(Z^a_{22}+Z^b_{11})^{\dagger}Z^b_{12},\\
Z^c_{21}&=Z^b_{21}(Z^a_{22}+Z^b_{11})^{\dagger}Z^a_{21},\\
Z^c_{22}&=Z^b_{22}-Z^b_{21}(Z^a_{22}+Z^b_{11})^{\dagger}Z^b_{12}.\\
\end{aligned}
\end{equation}

A simple but frequently-used cascade connection is the \textit{cascade-load} connection when $t=0$; hence, network $B$ can be treated as a load. 
The reduced expression of $Z^c$ is as follows:
\begin{equation}\label{eq:cascade-load Zc}
Z^c=Z^a_{11}-Z^a_{12}(Z^a_{22}+Z^b_{11})^{\dagger}Z^a_{21}.
\end{equation}
Clearly, the shorted connection is a special representation of the cascade-load connection. Series and parallel connections can also be placed under this category, as addressed in \cite{anderson2013network}.

\textit{Hybrid-Cascade connection:}
From Fig.~\ref{fig:hybrid-cascade connection}, we
partitioned $Z^a, ~Z^b$, and $Z^c$ as
\begin{equation}
\begin{aligned}
Z^a&\in \mathcal{R}^{\left( r+s\right) \times \left( r+s\right) },
\\
Z^b&\in \mathcal{R}^{\left( s+r\right) \times \left( s+r\right) },
\end{aligned}
\nonumber
\end{equation}
and
\begin{equation}
\begin{aligned}
~~~~Z^c&\in \mathcal{R}^{\left( r+s\right) \times \left( r+s\right) },
\end{aligned}
\nonumber
\end{equation}
wherein
\begin{equation}\label{eq:Hybrid-Cascade Zc2}
\begin{aligned}
Z^c_{11}&=Z^a_{11}+Z^b_{22}-(Z^a_{12}-Z^b_{21})(Z^a_{22}+Z^b_{11})^{\dagger}(Z^a_{21}-Z^b_{12}),\\
Z^c_{12}&=Z^b_{22}+(Z^a_{12}-Z^b_{21})(Z^a_{22}+Z^b_{11})^{\dagger}Z^b_{12},\\
Z^c_{21}&=Z^b_{22}+Z^b_{21}(Z^a_{22}+Z^b_{11})^{\dagger}(Z^a_{21}-Z^b_{12}),\\
Z^c_{22}&=Z^b_{22}-Z^b_{21}(Z^a_{22}+Z^b_{11})^{\dagger}Z^b_{12}.\\
\end{aligned}
\end{equation}

\noindent
\textbf{Subtractions}

The block dimensions of $Z^b$ and $Z^c$ under the subtractions given below remain unchanged from those of 
$Z^b$ and $Z^c$, respectively, under these connections. Therefore, the block dimensions of $Z^x$
can be determined according to the subsequent explicit formulae.

\textit{Hybrid subtraction:}
We have $Z^x\in \mathcal{R}^{\left( r+(n-r)\right) \times \left( r+(n-r)\right) }$,
where
\begin{equation}\label{eq:Hybrid Zx2}
\begin{aligned}
Z^x_{11}&=Z^c_{11}-Z^b_{11}-(Z^c_{12}-Z^b_{12})(Z^c_{22}-Z^b_{22})^{-1}(Z^c_{21}-Z^b_{21}),\\
Z^x_{12}&=Z^b_{12}-(Z^c_{12}-Z^b_{12})(Z^c_{22}-Z^b_{22})^{-1}Z^b_{22},\\
Z^x_{21}&=Z^b_{21}-Z^b_{22}(Z^c_{22}-Z^b_{22})^{-1}(Z^c_{21}-Z^b_{21}),\\
Z^x_{22}&=-Z^b_{22}(Z^c_{22}-Z^b_{22})^{-1}Z^c_{22}.
\end{aligned}
\end{equation}

\textit{Cascade subtraction:}
It follows from $Z^x\in \mathcal{R}^{\left( r+s\right) \times \left( r+s\right) }$ that:
\begin{equation}\label{eq:Cascade Zx2}
\begin{aligned}
Z^x_{11}&=Z^c_{11}-Z^c_{12}(Z^c_{22}-Z^b_{22})^{-1}Z^c_{21},\\
Z^x_{12}&=-Z^c_{12}(Z^c_{22}-Z^b_{22})^{-1}Z^b_{21},\\
Z^x_{21}&=-Z^b_{12}(Z^c_{22}-Z^b_{22})^{-1}Z^c_{21},\\
Z^x_{22}&=-Z^b_{11}-Z^b_{12}(Z^c_{22}-Z^b_{22})^{-1}Z^b_{21}.\\
\end{aligned}
\end{equation}

\textit{Hybrid-Cascade subtraction:}
The blocks of $Z^x\in \mathcal{R}^{\left( r+s\right) \times \left( r+s\right) }$ are expressed as:
\begin{equation}\label{eq:Hybrid-Cascade Zx2}
\begin{aligned}
Z^x_{11}&=Z^c_{11}-Z^b_{22}-(Z^c_{12}-Z^b_{22})(Z^c_{22}-Z^b_{11})^{-1}(Z^c_{21}-Z^b_{22}),\\
Z^x_{12}&=Z^b_{21}-(Z^c_{12}-Z^b_{22})(Z^c_{22}-Z^b_{22})^{-1}Z^b_{21},\\
Z^x_{21}&=Z^b_{12}-Z^b_{12}(Z^c_{22}-Z^b_{22})^{-1}(Z^c_{21}-Z^b_{22}),\\
Z^x_{22}&=-Z^b_{11}-Z^b_{12}(Z^c_{22}-Z^b_{22})^{-1}Z^b_{21}.\\
\end{aligned}
\end{equation}

\section*{Appendix B}
To prove Theorem \ref{thm:phase-preserving}, we first require the following auxiliary lemmas.

\begin{lemma} ($\!$\cite{wang2020phases}) \label{lemma:sum}
If $Z^a, Z^b\in \mathcal{Z}[\alpha, \beta],$ then
$Z^c=Z^a+Z^b\in \mathcal{Z}[\alpha, \beta].$
\end{lemma}

\begin{lemma} ($\!$\cite{chen2021phase}) \label{lemma:pseudo inverse}
If $Z^a\in \mathcal{Z}[\alpha, \beta],$ then 
$Z^{a\dagger}\in \mathcal{Z}[-\beta -\alpha].$
\end{lemma}

\begin{lemma} ($\!$\cite{furtado2003spectral}) \label{lemma:submatrix}
If $Z^a\in \mathcal{Z}[\alpha,\beta],$ then principal submatrices 
$Z^a_{ii}\in \mathcal{Z}[\alpha, \beta].$
\end{lemma}

\begin{lemma}($\!$\cite{wang2020phases}) \label{lemma:congruence}
If $Z^a\in \mathcal{Z}[\alpha, \beta],$ then for any nonsingular matrix $P$, the congruent transformation
$P^{\ast}Z^{a}P\in \mathcal{Z}[\alpha, \beta].$
\end{lemma}

\begin{lemma} \label{lemma:pseudo schur}
If $Z^a\in \mathcal{Z}[\alpha, \beta],$ then its  Schur complement
$Z^a/_{22}=Z_{11}^a-Z_{12}^aZ_{22}^{a\dagger}Z_{21}^a \in \mathcal{Z}[\alpha, \beta].$
\end{lemma}
\medskip\noindent {\em Proof of Lemma \ref{lemma:pseudo schur}:}
For a Z-matrix 
$$Z^a=\begin{bmatrix}
Z_{11}^a&Z_{12}^a
\\
Z_{21}^a&Z_{22}^a
\end{bmatrix},$$
it shall satisfy the existence condition  \eqref{eq:well-posedness}. Then it follows from~\cite{burns1974generalized} that the Moore-Penrose inverse can be partitioned as
$$Z^{a\dagger}=\begin{bmatrix}
Z^{a\dagger} _{11}& Z^{a\dagger} _{12}
\\
Z^{a\dagger}_{21}&Z^{a\dagger} _{22}
\end{bmatrix},$$
where $Z^{a\dagger} _{11}=\left(Z_{11}^a-Z_{12}^a Z_{22}^{a\dagger}Z_{21}^a \right)^{\dagger}\!\!. $
Hence, based on Lemma~\ref{lemma:pseudo inverse} and Lemma~\ref{lemma:submatrix}, we obtain $Z^{a\dagger}\in \mathcal{Z}[-\beta, -\alpha],$ such that $ Z^{a\dagger}_{11}$, which yields the lemma.
\hfill $\blacksquare$

Now we are ready to prove Theorem \ref{thm:phase-preserving}.

\textit{Proof of Theorem \ref{thm:phase-preserving}:}
We examined the connections \textbf{(i)–(vii)} individually.

\noindent (i) Shorted connection:

\noindent The phase-preserving result \eqref{eq:phase-preserving 1} can be established by Lemma~\ref{lemma:pseudo schur} directly.

\noindent (ii) Open connection:

\noindent The result \eqref{eq:phase-preserving 2} follows from Lemma~\ref {lemma:submatrix} immediately.

\noindent (iii) Series connection:

\noindent Lemma~\ref{lemma:sum} helps us to arrive at the result.

\noindent (iv) Parallel connection:

\noindent Owing to the well-posed nature assumption \eqref{eq:well-posedness}, it has been shown in \cite{mitra1986parallel}
 that 
$$Z^c=Z^{a}\left(Z^{a}+Z^{b}\right)^\dagger Z^{b}=\left( Z^{a\dagger}+Z^{b\dagger}\right) ^{\dagger}.$$
From Lemmas \ref{lemma:sum} and \ref{lemma:pseudo inverse}, we have $Z^c\in \mathcal{Z}[\alpha, \beta]$.

\noindent (v) Hybrid connection:

\noindent As $Z^a, Z^b\in \mathcal{Z}[\alpha, \beta]$, through filling zeros,
we have augmented Z matrices $\hat{Z}^a, \hat{Z}^b\in \mathcal{Z}[\alpha, \beta],$
where 
\begin{equation}\label{eq:augmented Za}
\hat{Z}^a=\begin{bmatrix}
Z_{11}^a&Z_{12}^a&0
\\
Z_{21}^a&Z_{22}^a&0
\\
0&0&0
\end{bmatrix}~\mathrm{and}~~\hat{Z}^b=\begin{bmatrix}
0&0&0
\\
0&Z_{11}^b&Z_{12}^b
\\
0&Z_{21}^b&Z_{22}^b

\end{bmatrix}.
\end{equation}
Multiplying the left side of $\hat{Z}^a$ and $\hat{Z}^b$ by the nonsingular matrices $T^{a \ast}$ and $T^{b \ast}$
and the right-hand side by $T^{a}$ and $T^{b}$, respectively, we obtain 
$$\bar{Z}^a=T^{a \ast}\hat{Z}^aT^{a}=\begin{bmatrix}
Z_{11}^a &Z_{12}^a &Z_{12}^a
\\
Z_{21}^a &Z_{22}^a &Z_{22}^a
\\
Z_{21}^a &Z_{22}^a &Z_{22}^a
\end{bmatrix},$$
and
$$\bar{Z}^b=T^{b \ast}\hat{Z}^bT^{b}=\begin{bmatrix}
Z_{11}^b &0&-Z_{12}^b 
\\
0&0&0
\\
-Z_{21}^b &0&Z_{22}^b

\end{bmatrix},$$
where
$$T^{a}=\begin{bmatrix}
I&0&0
\\
0&I&I
\\
0&0&-I
\end{bmatrix}
~\mathrm{and}~ 
T^{b}=\begin{bmatrix}
0&I&0
\\
-I&0&0
\\
0&0&I
\end{bmatrix}.
$$
According to Lemma~\ref{lemma:congruence}, we can claim that
$$\bar{Z}^a, \bar{Z}^b\in \mathcal{Z}[\alpha, \beta].$$
Furthermore, from Lemma~\ref{lemma:sum}, it follows that $\bar{Z}^c=\bar{Z}^a+\bar{Z}^b\in \mathcal{Z}[\alpha, \beta]$, where $\bar{Z}^c=$
$$\begin{bmatrix}
Z_{11}^a+Z_{11}^b&Z_{12}^a&Z_{12}^a-Z_{12}^b
\\
Z_{21}^a&Z_{22}^a&Z_{22}^a
\\
Z_{21}^a-Z_{21}^b&Z_{22}^a&Z_{22}^a+Z_{22}^b
\end{bmatrix}.$$
Based on Lemma~\ref{lemma:pseudo schur}, we obtain
$$
\begin{aligned}
&~~\bar{Z}^c/_{22}
\\
&\!=\!\begin{bmatrix}
Z_{11}^a+Z_{11}^b&Z_{12}^a
\\
Z_{21}^a&Z_{22}^a
\end{bmatrix}
\!-\!\begin{bmatrix}
Z_{12}^a-Z_{12}^b
\\
Z_{22}^a
\end{bmatrix} \left( Z_{22}^a+Z_{22}^b\right)^{\dagger} \times
\\
&~~~~~~~~~~~~~~~~~~~~~~~~~~~~~~~~~~~~~
\begin{bmatrix}
Z_{21}^a-Z_{21}^b & Z_{22}^a
\end{bmatrix}
\\
&=\begin{bmatrix}
(\bar{Z}^c/_{22})_{11}&(\bar{Z}^c/_{22})_{12}
\\
(\bar{Z}^c/_{22})_{21}&(\bar{Z}^c/_{22})_{22}
\end{bmatrix}
\in \mathcal{Z}[\alpha, \beta], 
\end{aligned}
$$
where
$(\bar{Z}^c/_{22})_{11}=Z^c_{11}$, $(\bar{Z}^c/_{22})_{12}=Z^c_{12}$, $(\bar{Z}^c/_{22})_{21}=Z^c_{21}$ in \eqref{eq:Hybrid Zc2}, and
$(\bar{Z}^c/_{22})_{22}=Z^a_{22}-Z^a_{22}(Z^a_{22}+Z^b_{22})^{\dagger}Z^a_{22}$.

It was proven in \cite{burns1974generalized}, that based on the condition of well-posed nature \eqref{eq:well-posedness} to the matrices $Z^a$ and $Z^b$, $Z^a_{22}(Z^a_{22}+Z^b_{22})^{\dagger}(Z^a_{22}+Z^b_{22})=Z^a_{22}$, which leads to the equality
$$Z^a_{22}-Z^a_{22}(Z^a_{22}+Z^b_{22})^{\dagger}Z^a_{22}=Z^a_{22}(Z^a_{22}+Z^b_{22})^{\dagger}Z^b_{22}.$$
Therefore, $(\bar{Z}^c/_{22})_{22}$ coincides with $Z^c_{22}$ in \eqref{eq:Hybrid Zc2}. Finally, for $Z^c$ in \eqref{eq:Hybrid Zc2}, 
we obtain $Z^c \in \mathcal{Z}[\alpha, \beta].$

Regarding cascade and hybrid-cascade connections, we can follow the procedures outlined above to obtain the phase preservation result \eqref{eq:phase-preserving 3}. Additionally, in Section \ref{sec:Extension}, we will present a more comprehensive result with complete proof. Therefore, we omit the proof for the remaining two connections for the sake of brevity.
\hfill $\blacksquare$

\section*{Appendix C}
\textit{Proof of Theorem \ref{thm:phase-preserving2}:}
For $Z^{b} \in \mathcal{Z}[\alpha^{b}, \beta^{b}]$, 
it follows that $-Z^{b}\in\mathcal{Z}[\alpha^{b}+\pi , \beta^{b}+\pi ]$ mod  $2\pi$.
Meanwhile, if $[\alpha^{b}, \beta^{b}]\cap [\alpha^{c}, \beta^{c}]=\emptyset$ mod $2\pi$, we can easily verify that 
$\max\lbrace   \beta^{c},   \beta^{b}+\pi  \rbrace-\min\lbrace   \alpha^{c},   \alpha^{b}+\pi    \rbrace<\pi$ or
$\max\lbrace   \beta^{c},   \beta^{b}-\pi  \rbrace-\min\lbrace   \alpha^{c},   \alpha^{b}-\pi    \rbrace<\pi$.
Hence, without loss of generality, we assume that the former holds. This gives rise to
$$ -Z^{b}\in\mathcal{Z}[\min\lbrace \alpha^{c}, \alpha^{b}+\pi \rbrace,    \max\lbrace   \beta^{c},   \beta^{b}+\pi  \rbrace  ].$$
In addition,  we have 
$$Z^{c}\in  \mathcal{Z}[\alpha^{c}, \beta^{c}] \subseteq \mathcal{Z}[\min\lbrace   \alpha^{c},   \alpha^{b}+\pi  \rbrace,    \max\lbrace   \beta^{c},   \beta^{b}+\pi  \rbrace  ].$$
The remainder of the proof is analogous to that of Theorem~\ref{thm:phase-preserving}; for this reason, we provide an outline.

\noindent (i) Series subtraction and parallel subtraction:

The proofs are similar to those of series and parallel connections and obtained based on Lemmas \ref{lemma:sum} and \ref{lemma:pseudo inverse}.

\noindent (ii) Hybrid subtraction:

Let $\bar{Z}^c=T^{c \ast}\hat{Z}^cT^{c}$ and $\bar{Z}^b=T^{b \ast}\hat{Z}^bT^{b}$,
where 
\begin{equation}\label{eq:subtraction:Zc}
\hat{Z}^c=\begin{bmatrix}
0&0&0
\\
0&Z_{11}^c&Z_{12}^c
\\
0&Z_{21}^c&Z_{22}^c
\end{bmatrix},~~
\hat{Z}^b=\begin{bmatrix}
Z_{11}^b&Z_{12}^b&0
\\
Z_{21}^b&Z_{22}^b&0
\\
0&0&0
\end{bmatrix},
\end{equation}
and
\begin{equation}
T^{c}=\begin{bmatrix}
0&I&0
\\
I&0&0
\\
0&0&I
\end{bmatrix},
~~
T^{b}=\begin{bmatrix}
I&0&0
\\
0&-I&I
\\
0&0&-I
\end{bmatrix}.
\end{equation}
Then, it follows that
$$
\begin{aligned}
\bar{Z}^x&=\bar{Z}^c-\bar{Z}^b
\\
&=
\begin{bmatrix}
Z_{11}^c-Z_{11}^b& Z_{12}^b&Z_{12}^c-Z_{12}^b
\\
Z_{21}^b& -Z_{22}^b&Z_{22}^b
\\
Z_{21}^c-Z_{21}^b& Z_{22}^b&Z_{22}^c-Z_{22}^b
\end{bmatrix}.
\end{aligned}
$$
Based on Lemmas \ref{lemma:sum} and \ref{lemma:congruence}, we can claim that 
$$\bar{Z}^x\in \mathcal{Z}[\min\lbrace   \alpha^{c},   \alpha^{b}+\pi  \rbrace, \max\lbrace \beta^{c}, \beta^{b}+\pi  \rbrace].$$
Considering the Schur complement of $\bar{Z}^x$ with respect to $Z_{22}^c-Z_{22}^b$, we obtain
$Z^x$ in Eq. \eqref{eq:Hybrid Zx1}. Hence,  
using Lemma~\ref{lemma:pseudo schur}, we obtain the result \eqref{eq:thm2 condition} for hybrid subtraction.

The proof details for cascade and hybrid-cascade subtractions can be addressed similarly to those of hybrid subtraction. We omit the remaining procedures.
\hfill $\blacksquare$

\section*{Appendix D}
\textit{Proof of Theorem \ref{thm: confluence representation}:}
Given a feasible matrix representation $\stbt{\Phi}{\Psi}{\Xi}{\Omega}$ to $\textbf{G}^{\perp}$, the explicit formula $Z^c=M/_{22}$ in \eqref{eq:Zc}
has been given in~\cite{anderson1975matrix}. Recalling the condition \eqref{eq:well-posedness},
 the existence condition of $Z^c$ can be obtained accordingly as follows:
 \begin{equation}\label{eq:Zc existence condition1}
\mathcal{R}\left( \Xi Z^a \Phi^{\ast}+\Omega Z^b \Psi^{\ast}\right) \subseteq \mathcal{R}\left( \Xi Z^a \Xi^{\ast}+\Omega Z^b \Omega^{\ast}\right),
\end{equation}
and
\begin{equation}\label{eq:Zc existence condition2}
\mathcal{N}\left( \Phi Z^a \Xi^{\ast}+\Psi Z^b \Omega^{\ast}\right) \supseteq \mathcal{N}\left( \Xi Z^a \Xi^{\ast}+\Omega Z^b \Omega^{\ast}\right).
\end{equation}

We next show that the formula  \eqref{eq:Zc} and the existence condition \eqref{eq:Zc existence condition1}-\eqref{eq:Zc existence condition2} are valid for all possible 
matrix representations $\stbt{\Phi}{\Psi}{\Xi}{\Omega}$ to $\textbf{G}^{\perp}$. That is their expressions are independent of the parameter matrices  $\Gamma$ and $\Lambda$ defined in  \eqref{eq:confluence perp parameterization}.  


Let $\Phi+\Gamma \Xi$, $\Psi+\Gamma\Omega$, $\Lambda\Xi$, and $\Lambda\Omega$ replace $\Phi$, $\Psi$, $\Xi$, and $\Omega$, as in \eqref{eq:confluence perp parameterization}.
Then we show that  $\Gamma$ and $\Lambda$ are not involved in the existence condition \eqref{eq:Zc existence condition1}.
The corresponding condition involving $\Gamma$ and $\Lambda$ is given by
\begin{equation}\label{eq:existence Zc with Gamma and Lambda}
\begin{aligned}
&\mathcal{R}\left( \Lambda\Xi Z^a (\Phi+\Gamma \Xi)^{\ast}+\Lambda\Omega Z^b (\Psi+\Gamma\Omega)^{\ast}\right) 
\\
&\qquad \qquad \subseteq \mathcal{R}\left( \Lambda\Xi Z^a (\Lambda\Xi)^{\ast}+\Lambda\Omega Z^b (\Lambda\Omega)^{\ast}\right).
\end{aligned}
\end{equation}
Condition \eqref{eq:Zc existence condition1} $\Rightarrow$ \eqref{eq:existence Zc with Gamma and Lambda}:
Note that in \eqref{eq:existence Zc with Gamma and Lambda}, 
$$
\begin{aligned}
&\Lambda\Xi Z^a (\Phi+\Gamma \Xi)^{\ast}+\Lambda\Omega Z^b (\Psi+\Gamma\Omega)^{\ast}
\\
&~=
\Lambda\Xi Z^a\Phi^{\ast}+\Lambda\Omega Z^b\Psi^{\ast}+\Lambda\Xi Z^a \Xi^{\ast}\Gamma^{\ast} +\Lambda\Omega Z^b\Omega^{\ast} \Gamma^{\ast}.
\end{aligned}
$$
We also have
\begin{equation}\label{eq:existence Zc with Gamma and Lambda2}
\begin{aligned}
\mathcal{R}\left( \Lambda\Xi Z^a \Xi^{\ast}\Gamma^{\ast} +\Lambda\Omega Z^b\Omega^{\ast} \Gamma^{\ast}\right) \subseteq \mathcal{R}\left( \Lambda\Xi Z^a \Xi^{\ast}+\Lambda\Omega Z^b\Omega^{\ast} \right)
\\
\subseteq \mathcal{R}\left( \Lambda\Xi Z^a \Xi^{\ast}\Lambda^{\ast}+\Lambda\Omega Z^b\Omega^{\ast}\Lambda^{\ast} \right)
\end{aligned}
\end{equation}
since $\mathrm{Rank}(\Lambda)=n$.
Therefore, the relation in \eqref{eq:existence Zc with Gamma and Lambda} holds if we have
\begin{equation}\label{eq:existence Zc with Gamma and Lambda3}
\begin{aligned}
\mathcal{R}\left(\Lambda\Xi Z^a\Phi^{\ast}+\Lambda\Omega Z^b\Psi^{\ast} \right)
\subseteq \mathcal{R}\left( \Lambda\Xi Z^a \Xi^{\ast}\Lambda^{\ast}+\Lambda\Omega Z^b\Omega^{\ast}\Lambda^{\ast} \right),
\end{aligned}
\end{equation}
which can be implied by the condition \eqref{eq:Zc existence condition1} clearly.

\noindent Condition \eqref{eq:existence Zc with Gamma and Lambda} $\Rightarrow$ \eqref{eq:Zc existence condition1}:
Note that the condition \eqref{eq:existence Zc with Gamma and Lambda2} always holds. Therefore, the condition \eqref{eq:existence Zc with Gamma and Lambda} yields \eqref{eq:existence Zc with Gamma and Lambda3},
so is to \eqref{eq:Zc existence condition1} since $\Lambda$ is full rank.

Another existence condition \eqref{eq:Zc existence condition2} can be verified analogously and hence omitted herein.

Finally, we prove that  $Z^c=M/_{22}$ is also independent of  $\Gamma$ and $\Lambda$.
Denote by 
$\Pi=\stbt{I}{\Gamma}{0}{\Lambda}.$ Partition $M$ in \eqref{eq:Zc general} compatibly as $$
\begin{bmatrix}
M_{11}&M_{12}
\\
M_{21}&M_{22}
\end{bmatrix}.
$$
It suffices to show that $M/_{22}=(\Pi M\Pi^{\ast})/_{22}.$
In turn, we  have 
\begin{equation}
\begin{aligned}
&(\Pi M\Pi^{\ast})/_{22}
\\
&=M_{11}+\Gamma M_{21}+M_{12} \Gamma^{\ast} +\Gamma M_{22} \Gamma^{\ast}
\\
&~~~~
-\left( M_{12}\Lambda^{\ast}+\Gamma M_{22} \Lambda^{\ast}\right) \left( \Lambda M_{22} \Lambda^{\ast} \right)^{\dagger} \left( \Lambda M_{21}+\Lambda M_{22}\Gamma^{\ast}\right) 
\\
&=M_{11}+\Gamma M_{21}+M_{12} \Gamma^{\ast} +\Gamma M_{22} \Gamma^{\ast}
\\
&~~~~
-\left( M_{12}+\Gamma M_{22}\right) M_{22}^{\dagger} \left(  M_{21}+M_{22}\Gamma^{\ast}\right) 
\\
&=M/_{22}+\Gamma M_{21}+M_{12} \Gamma^{\ast} +\Gamma M_{22} \Gamma^{\ast}-M_{12}M_{22}^{\dagger}M_{22}\Gamma^{\ast}
\\
&~~~~
-\Gamma M_{22}M_{22}^{\dagger}M_{21}-\Gamma M_{22} M_{22}^{\dagger}M_{22}\Gamma^{\ast}
\\
&=M/_{22}.
\end{aligned}
\nonumber
\end{equation}
Note that as investigated in \cite{burns1974generalized}, the equalities $M_{12}M_{22}^{\dagger}M_{22}=M_{12}$, 
$M_{22}M_{22}^{\dagger}M_{21}=M_{21}$, and $M_{22} M_{22}^{\dagger}M_{22}=M_{22}$ can be established as long as the
existence condition of $Z^c$ in \eqref{eq:Zc existence condition1}-\eqref{eq:Zc existence condition2} holds.
 \hfill $\blacksquare$

\section*{Appendix E}
First, a new supporting lemma is provided as follows.
\begin{lemma} ($\!\!$\cite{wang2020phases}) \label{lemma:compression}
If $Z^a\in \mathcal{Z}[\alpha^a, \beta^a]$, then for all full row rank matrix $J\in \mathbb{C}^{(n-k)\times n}$, we have $JZ^aJ^{\ast}\in \mathcal{Z}[\alpha^a, \beta^a]$.
\end{lemma}

\textit{Proof of Theorem \ref{thm:phase-preserving confluence}:}
As a direct consequence of Lemma~\ref{lemma:pseudo schur},
for $Z^a, Z^b\in \mathcal{Z}[\alpha, \beta],$ 
we have $Z^c=M/_{22} \in \mathcal{Z}[\alpha, \beta]$ if $M \in \mathcal{Z}[\alpha, \beta]$.
Clearly, if 
$$
\mathrm{Rank}\left( \begin{bmatrix}
\Phi&\Psi\\
\Xi&\Omega
\end{bmatrix}  \right)=2n, 
$$
it follows from  Lemma \ref{lemma:congruence} that
\begin{flalign}
M=\begin{bmatrix}
\Phi&\Psi\\
\Xi&\Omega
\end{bmatrix} 
\begin{bmatrix}
Z^a&0\\
0&Z^b
\end{bmatrix} 
\begin{bmatrix}
\Phi^{\ast}&\Xi^{\ast}\\
\Psi^{\ast}&\Omega^{\ast}
\end{bmatrix}\in \mathcal{Z}[\alpha, \beta].
\nonumber
\end{flalign}
If
$$
\mathrm{Rank}\left( \begin{bmatrix}
\Phi&\Psi\\
\Xi&\Omega
\end{bmatrix}  \right)=k<2n, 
$$
we can always decompose the matrix as
$\stbt{\Phi}{\Psi}{\Xi}{\Omega}=XY$,
where $X,~Y\in \mathbb{R}^{2n\times 2n}$, $\mathrm{Rank}(X)=2n$, and  $\mathrm{Rank}(Y)=k$. More specifically,
the matrix $Y$ takes the form of 
$$Y=\begin{bmatrix}
\Sigma
\\
0
\end{bmatrix},$$
where $\Sigma\in \mathbb{R}^{k\times 2n}.$ Substituting $X$ and $Y$ into the expression of $M$, we obtain that
\begin{equation}
\begin{aligned}
M&=XY 
\begin{bmatrix}
Z^a&0\\
0&Z^b
\end{bmatrix} 
Y^{\ast}X^{\ast}
\\
&=X\begin{bmatrix}
\Sigma
\\
0
\end{bmatrix}
\begin{bmatrix}
Z^a&0\\
0&Z^b
\end{bmatrix} 
\begin{bmatrix}
\Sigma^{\ast}
&
0
\end{bmatrix}X^{\ast}
\\
&=X\begin{bmatrix}
\Sigma \left[ \begin{smallmatrix}
Z^a&0
\\
0&Z^b
\end{smallmatrix}\right] 
 \Sigma^{\ast}&0\\
0&0
\end{bmatrix}X^{\ast}.
\end{aligned}
\nonumber
\end{equation}
Since $\Sigma$ is full row rank, Lemma~\ref{lemma:compression} points out that
$\Sigma \left[ \begin{smallmatrix}
Z^a&0
\\
0&0
\end{smallmatrix}\right] 
 \Sigma^{\ast}\in \mathcal{Z}[\alpha^a, \beta^a].$
The proof hence is completed by invoking Lemma~\ref{lemma:sum} owing to the full rank matrix $X$.
\hfill $\blacksquare$

\bibliographystyle{IEEEtran}

\bibliography{mycite}

\begin{thebibliography}{10}
\providecommand{\url}[1]{#1}
\csname url@samestyle\endcsname
\providecommand{\newblock}{\relax}
\providecommand{\bibinfo}[2]{#2}
\providecommand{\BIBentrySTDinterwordspacing}{\spaceskip=0pt\relax}
\providecommand{\BIBentryALTinterwordstretchfactor}{4}
\providecommand{\BIBentryALTinterwordspacing}{\spaceskip=\fontdimen2\font plus
\BIBentryALTinterwordstretchfactor\fontdimen3\font minus
  \fontdimen4\font\relax}
\providecommand{\BIBforeignlanguage}[2]{{%
\expandafter\ifx\csname l@#1\endcsname\relax
\typeout{** WARNING: IEEEtran.bst: No hyphenation pattern has been}%
\typeout{** loaded for the language `#1'. Using the pattern for}%
\typeout{** the default language instead.}%
\else
\language=\csname l@#1\endcsname
\fi
#2}}
\providecommand{\BIBdecl}{\relax}
\BIBdecl

\bibitem{desoer1966basic}
C.~A. Desoer and E.~S. Kuh, \emph{Basic Circuit Theory}.\hskip 1em plus 0.5em
  minus 0.4em\relax New York, NY: McGraw-Hill, 1966.

\bibitem{anderson2013network}
B.~D. Anderson and S.~Vongpanitlerd, \emph{Network Analysis and Synthesis: A
  Modern Systems Theory Approach}.\hskip 1em plus 0.5em minus 0.4em\relax
  Englwood Cliffs, NJ: Prentice-Hall, 2013.

\bibitem{laib2023decentralized}
K.~Laib, J.~Watson, Y.~Ojo, and I.~Lestas, ``Decentralized stability conditions
  for {DC} microgrids: Beyond passivity approaches,'' \emph{Automatica}, vol.
  149, p. 110705, 2023.

\bibitem{smith2000performance}
M.~C. Smith and G.~W. Walker, ``Performance limitations and constraints for
  active and passive suspensions: {A} mechanical multi-port approach,''
  \emph{Vehicle System Dynamics}, vol.~33, no.~3, pp. 137--168, 2000.

\bibitem{smith2002synthesis}
M.~C. Smith, ``Synthesis of mechanical networks: {T}he inerter,'' \emph{IEEE
  Transactions on Automatic Control}, vol.~47, no.~10, pp. 1648--1662, 2002.

\bibitem{chen2009missing}
M.~Z. Chen, C.~Papageorgiou, F.~Scheibe, F.-C. Wang, and M.~C. Smith, ``The
  missing mechanical circuit element,'' \emph{IEEE Circuits and Systems
  Magazine}, vol.~9, no.~1, pp. 10--26, 2009.

\bibitem{shearer1967introduction}
J.~L. Shearer, A.~T. Murphy, and H.~H. Richardson, \emph{Introduction to System
  Dynamics}.\hskip 1em plus 0.5em minus 0.4em\relax Addison-Wesley, 1967.

\bibitem{perelson1975network}
A.~S. Perelson, ``Network thermodynamics: {A}n overview.'' \emph{Biophysical
  Journal}, vol.~15, no.~7, p. 667, 1975.

\bibitem{youla2015theory}
D.~C. Youla, \emph{Theory and Synthesis of Linear Passive Time-Invariant
  Networks}.\hskip 1em plus 0.5em minus 0.4em\relax Cambridge, UK: Cambridge
  University Press, 2015.

\bibitem{shannon1949synthesis}
C.~E. Shannon, ``The synthesis of two-terminal switching circuits,'' \emph{The
  Bell System Technical Journal}, vol.~28, no.~1, pp. 59--98, 1949.

\bibitem{biorci1961synthesis}
G.~Biorci and P.~Civalleri, ``On the synthesis of resistive $n$-port
  networks,'' \emph{IRE Transactions on Circuit Theory}, vol.~8, no.~1, pp.
  22--28, 1961.

\bibitem{carlin1964singular}
H.~Carlin, ``Singular network elements,'' \emph{IEEE Transactions on Circuit
  Theory}, vol.~11, no.~1, pp. 67--72, 1964.

\bibitem{pozar2011microwave}
D.~M. Pozar, \emph{Microwave Engineering}.\hskip 1em plus 0.5em minus
  0.4em\relax New York, NY: John Wiley \& Sons, 2011.

\bibitem{odabasioglu1998prima}
A.~Odabasioglu, M.~Celik, and L.~T. Pileggi, ``Prima: {P}assive reduced-order
  interconnect macromodeling algorithm,'' \emph{IEEE Transactions on
  Computer-Aided Design of Integrated Circuits and Systems}, vol.~17, no.~8,
  pp. 645--654, 1998.

\bibitem{ortega2003power}
R.~Ortega, D.~Jeltsema, and J.~M. Scherpen, ``Power shaping: {A} new paradigm
  for stabilization of nonlinear {RLC} circuits,'' \emph{IEEE Transactions on
  Automatic Control}, vol.~48, no.~10, pp. 1762--1767, 2003.

\bibitem{garcia2007power}
E.~Garcia-Canseco, R.~Grino, R.~Ortega, M.~Salichs, and A.~M. Stankovic,
  ``Power-factor compensation of electrical circuits,'' \emph{IEEE Control
  Systems Magazine}, vol.~27, no.~2, pp. 46--59, 2007.

\bibitem{fiaz2013port}
S.~Fiaz, D.~Zonetti, R.~Ortega, J.~M. Scherpen, and A.~van~der Schaft, ``A
  port-{H}amiltonian approach to power network modeling and analysis,''
  \emph{European Journal of Control}, vol.~19, no.~6, pp. 477--485, 2013.

\bibitem{chaffey2021monotone}
T.~Chaffey and R.~J. Sepulchre, ``Monotone one-port circuits,'' \emph{IEEE
  Transactions on Automatic Control (Early Access)}, 2023.

\bibitem{miranda2022dissipativity}
F.~A. Miranda-Villatoro, F.~Forni, and R.~J. Sepulchre, ``Dissipativity
  analysis of negative resistance circuits,'' \emph{Automatica}, vol. 136, p.
  110011, 2022.

\bibitem{chaffey2023circuit}
T.~Chaffey, S.~Banert, P.~Giselsson, and R.~Pates, ``Circuit analysis using
  monotone+skew splitting,'' \emph{European Journal of Control}, p. 100854,
  2023.

\bibitem{van2000l2}
A.~van~der Schaft, \emph{$L_2$-Gain and Passivity Techniques in Nonlinear
  Control}.\hskip 1em plus 0.5em minus 0.4em\relax London, UK: Springer-Verlag,
  2000.

\bibitem{hill1980dissipative}
D.~J. Hill and P.~J. Moylan, ``Dissipative dynamical systems: {B}asic
  input-output and state properties,'' \emph{Journal of the Franklin
  Institute}, vol. 309, no.~5, pp. 327--357, 1980.

\bibitem{arcak2007passivity}
M.~Arcak, ``Passivity as a design tool for group coordination,'' \emph{IEEE
  Transactions on Automatic Control}, vol.~52, no.~8, pp. 1380--1390, 2007.

\bibitem{dorfler2014synchronization}
F.~D{\"o}rfler and F.~Bullo, ``Synchronization in complex networks of phase
  oscillators: A survey,'' \emph{Automatica}, vol.~50, no.~6, pp. 1539--1564,
  2014.

\bibitem{ren2010distributed}
W.~Ren and Y.~Cao, \emph{Distributed Coordination of Multi-Agent Networks:
  Emergent Problems, Models, and Issues}.\hskip 1em plus 0.5em minus
  0.4em\relax Springer Science \& Business Media, 2010.

\bibitem{li2020distributed}
M.~Li, L.~Su, and T.~Liu, ``Distributed optimization with event-triggered
  communication via input feedforward passivity,'' \emph{IEEE Control Systems
  Letters}, vol.~5, no.~1, pp. 283--288, 2020.

\bibitem{wang2020phases}
D.~Wang, W.~Chen, S.~Z. Khong, and L.~Qiu, ``On the phases of a complex
  matrix,'' \emph{Linear Algebra and Its Applications}, vol. 593, pp. 152--179,
  2020.

\bibitem{chen2021phase}
W.~Chen, D.~Wang, S.~Z. Khong, and L.~Qiu, ``A phase theory of {MIMO} {LTI}
  systems,'' \emph{arXiv preprint arXiv:2105.03630}, 2021.

\bibitem{mao2022phases}
X.~Mao, W.~Chen, and L.~Qiu, ``Phases of discrete-time {LTI} multivariable
  systems,'' \emph{Automatica}, vol. 142, p. 110311, 2022.

\bibitem{chen2020phase}
C.~Chen, D.~Zhao, W.~Chen, S.~Z. Khong, and L.~Qiu, ``Phase of nonlinear
  systems,'' \emph{arXiv preprint arXiv:2012.00692}, 2020.

\bibitem{Dan2023phase}
D.~Wang, W.~Chen, and L.~Qiu, ``Synchronization of diverse agents via phase
  analysis,'' \emph{Automatica (accepted)}, 2023.

\bibitem{postlethwaite1981principal}
I.~Postlethwaite, J.~Edmunds, and A.~MacFarlane, ``Principal gains and
  principal phases in the analysis of linear multivariable feedback systems,''
  \emph{IEEE Transactions on Automatic Control}, vol.~26, no.~1, pp. 32--46,
  1981.

\bibitem{reza1980concept}
F.~Reza, ``The concept of the phase of a linear $n$-port and a general
  dynamical system,'' \emph{Proceedings of the IEEE}, vol.~68, no.~4, pp.
  532--533, 1980.

\bibitem{freudenberg1988frequency}
J.~S. Freudenberg and D.~P. Looze, \emph{Frequency Domain Properties of Scalar
  and Multivariable Feedback Systems}.\hskip 1em plus 0.5em minus 0.4em\relax
  Heidelberg, Germany: Springer Berlin, 1988.

\bibitem{chen1998multivariable}
J.~Chen, ``Multivariable gain-phase and sensitivity integral relations and
  design trade-offs,'' \emph{IEEE Transactions on Automatic Control}, vol.~43,
  no.~3, pp. 373--385, 1998.

\bibitem{anderson1971shorted}
W.~N. Anderson, ``Shorted operators,'' \emph{SIAM Journal on Applied
  Mathematics}, vol.~20, no.~3, pp. 520--525, 1971.

\bibitem{anderson1975matrix}
W.~N. Anderson, R.~J. Duffin, and G.~E. Trapp, ``Matrix operations induced by
  network connections,'' \emph{SIAM Journal on Control}, vol.~13, no.~2, pp.
  446--461, 1975.

\bibitem{mitra1982shorted}
S.~K. Mitra and M.~L. Puri, ``Shorted matrices—an extended concept and some
  applications,'' \emph{Linear Algebra and Its Applications}, vol.~42, pp.
  57--79, 1982.

\bibitem{mitra1975hybrid}
S.~K. Mitra and G.~E. Trapp, ``On hybrid addition of matrices,'' \emph{Linear
  Algebra and Its Applications}, vol.~10, no.~1, pp. 19--35, 1975.

\bibitem{anderson1986cascade}
W.~N. Anderson, T.~D. Morley, and G.~E. Trapp, ``Cascade addition and
  subtraction of matrices,'' \emph{SIAM Journal on Algebraic Discrete Methods},
  vol.~7, no.~4, pp. 609--626, 1986.

\bibitem{mitra1986parallel}
S.~K. Mitra and P.~L. Odell, ``On parallel summability of matrices,''
  \emph{Linear Algebra and Its Applications}, vol.~74, pp. 239--255, 1986.

\bibitem{mitra1983fundamental}
S.~K. Mitra and M.~L. Puri, ``The fundamental bordered matrix of linear
  estimation and the {D}uffin-{M}orley general linear electromechanical
  systems,'' \emph{Applicable Analysis}, vol.~14, no.~4, pp. 241--258, 1983.

\bibitem{pekarev1992shorts}
E.~Pekarev, ``Shorts of operators and some extremal problems,'' \emph{Acta Sci.
  Math.(Szeged)}, vol.~56, no. 1-2, pp. 147--163, 1992.

\bibitem{antezana2006bilateral}
J.~Antezana, G.~Corach, and D.~Stojanoff, ``Bilateral shorted operators and
  parallel sums,'' \emph{Linear Algebra and Its Applications}, vol. 414, no.
  2-3, pp. 570--588, 2006.

\bibitem{zhao2019stabilization}
D.~Zhao, L.~Qiu, and G.~Gu, ``Stabilization of two-port networked systems with
  simultaneous uncertainties in plant, controller, and communication
  channels,'' \emph{IEEE Transactions on Automatic Control}, vol.~65, no.~3,
  pp. 1160--1175, 2019.

\bibitem{chen2020spectral}
W.~Chen, D.~Wang, J.~Liu, Y.~Chen, S.~Z. Khong, T.~Ba{\c{s}}ar, K.~H.
  Johansson, and L.~Qiu, ``On spectral properties of signed {L}aplacians with
  connections to eventual positivity,'' \emph{IEEE Transactions on Automatic
  Control}, vol.~66, no.~5, pp. 2177--2190, 2020.

\bibitem{maxwell1873treatise}
J.~C. Maxwell, \emph{A Treatise on Electricity and Magnetism}.\hskip 1em plus
  0.5em minus 0.4em\relax Oxford, UK: Clarendon Press, 1873.

\bibitem{chen2023port}
J.~Chen, W.~Chen, and L.~Qiu, ``Phase analysis of {N}-port electrical networks
  under interconnections,'' in \emph{Proceedings of the 22nd World Congress of
  the International Federation of Automatic Control (IFAC)}, Yokohama, Japan,
  2023.

\bibitem{zhang2015matrix}
F.~Zhang, ``A matrix decomposition and its applications,'' \emph{Linear and
  Multilinear Algebra}, vol.~63, no.~10, pp. 2033--2042, 2015.

\bibitem{roger1994topics}
H.~Roger and R.~J. Charles, \emph{Topics in Matrix Analysis}.\hskip 1em plus
  0.5em minus 0.4em\relax Cambridge, UK: Cambridge University Press, 1994.

\bibitem{furtado2003spectral}
S.~Furtado and C.~R. Johnson, ``Spectral variation under congruence for a
  nonsingular matrix with 0 on the boundary of its field of values,''
  \emph{Linear Algebra and Its Applications}, vol. 359, no. 1-3, pp. 67--78,
  2003.

\bibitem{qiu2022phases}
D.~Wang, X.~Mao, W.~Chen, and L.~Qiu, ``On the phases of a semi-sectorial
  matrix and the essential phase of a {L}aplacian,'' \emph{Linear Algebra and
  its Applications}, vol. 676, pp. 441--458, 2023.

\bibitem{georgiou2019principles}
T.~T. Georgiou, F.~Jabbari, and M.~C. Smith, ``Principles of lossless
  adjustable one-ports,'' \emph{IEEE Transactions on Automatic Control},
  vol.~65, no.~1, pp. 252--262, 2019.

\bibitem{smith2020inerter}
M.~C. Smith, ``The inerter: {A} retrospective,'' \emph{Annual Review of
  Control, Robotics, and Autonomous Systems}, vol.~3, pp. 361--391, 2020.

\bibitem{duffin1978almost}
R.~Duffin and T.~Morley, ``Almost definite operators and electro-mechanical
  systems,'' \emph{SIAM Journal on Applied Mathematics}, vol.~35, no.~1, pp.
  21--30, 1978.

\bibitem{anderson1975shorted}
W.~N. Anderson and G.~Trapp, ``Shorted operators. {II},'' \emph{SIAM Journal on
  Applied Mathematics}, vol.~28, no.~1, pp. 60--71, 1975.

\bibitem{burns1974generalized}
F.~Zhang, \emph{The Schur Complement And Its Applications}.\hskip 1em plus
  0.5em minus 0.4em\relax New York, NY: Springer Science \& Business Media,
  2006.

\bibitem{zhou1996robust}
K.~Zhou, J.~C. Doyle, and K.~Glover, \emph{Robust and Optimal Control}.\hskip
  1em plus 0.5em minus 0.4em\relax Englewood Cliffs, NJ: Prentice Hall, 1996.

\bibitem{anderson1969series}
W.~N. Anderson and R.~J. Duffin, ``Series and parallel addition of matrices,''
  \emph{Journal of Mathematical Analysis and Applications}, vol.~26, no.~3, pp.
  576--594, 1969.

\bibitem{steigerwald1988comparison}
R.~L. Steigerwald, ``A comparison of half-bridge resonant converter
  topologies,'' \emph{IEEE Transactions on Power Electronics}, vol.~3, no.~2,
  pp. 174--182, 1988.

\bibitem{willems1976realization}
J.~C. Willems, ``Realization of systems with internal passivity and symmetry
  constraints,'' \emph{Journal of the Franklin Institute}, vol. 301, no.~6, pp.
  605--621, 1976.

\bibitem{duffin1966network}
R.~J. Duffin, D.~Hazony, and N.~Morrison, ``Network synthesis through hybrid
  matrices,'' \emph{SIAM Journal on Applied Mathematics}, vol.~14, no.~2, pp.
  390--413, 1966.

\end{thebibliography}

\end{document}